\documentclass[a4,11pt]{article}

\usepackage[title]{appendix}
\usepackage{parskip,amsmath,amsfonts,amssymb,amsthm,graphicx,url,breqn,parcolumns,mathtools,multirow}
\usepackage{graphicx}
\usepackage[breakable,theorems]{tcolorbox}
\usepackage{xspace} 

\usepackage{algorithm}
\usepackage{algpseudocode}
\usepackage{setspace}
\floatname{algorithm}{}

\usepackage{euscript}
\usepackage{enumerate}
\usepackage{enumitem}
\setlist{leftmargin=5.5mm}
\usepackage{subfigure}

\usepackage{verbatim}

\algnewcommand{\IIf}[1]{\State\algorithmicif\ #1\ \algorithmicthen}
\algnewcommand{\EndIIf}{\unskip\ \algorithmicend\ \algorithmicif}

\algnewcommand{\IiIf}[1]{\State\algorithmicif\ #1\ \algorithmicthen}
\algnewcommand{\EndIiIf}{\unskip\ \algorithmicend\ \algorithmicif}

\algnewcommand{\IfThenElse}[3]{
 \State \algorithmicif\ #1\ \algorithmicthen\ #2\ \algorithmicelse\ #3}
\algnewcommand{\EndIfThenElse}{\unskip\ \algorithmicend\ \algorithmicif}

\newcommand{\Trees}{$SCNP_{tree}$\xspace}
\newcommand{\BDAL}{$BD_{SCNP}$\xspace}

\newcommand{\E}{\mathcal{E}}

\renewcommand{\O}{\mathcal{O}}
\renewcommand{\P}{\mathcal{P}}

\renewcommand{\S}{\mathbb{S}}
\newcommand{\T}{\mathcal{T}}

\newtcbtheorem[number within=section]{myproperty}{Property}{
width=\linewidth,
leftrule=1.5mm,bottomrule=0mm,toprule=0mm,rightrule=0mm,arc=0mm,
breakable,toprule at break=0mm,bottomrule at break=0mm,
colback=white,colframe=black,coltitle=black,colbacktitle=white,
fonttitle=\bfseries, 
}{th}
\newtcbtheorem[]{mycorollary}{Corollary}{
width=\linewidth,
leftrule=1.5mm,bottomrule=0mm,toprule=0mm,rightrule=0mm,arc=0mm,
breakable,toprule at break=0mm,bottomrule at break=0mm,
colback=white,colframe=black,coltitle=black,colbacktitle=white,
fonttitle=\bfseries, 
}{th}

\newtcbtheorem[]{myTheo}{Theorem}{
width=\linewidth,
leftrule=1.5mm,bottomrule=0mm,toprule=0mm,rightrule=0mm,arc=0mm,
breakable,toprule at break=0mm,bottomrule at break=0mm,
colback=white,colframe=black,coltitle=black,colbacktitle=white,
fonttitle=\bfseries, 
}{th}

\newtcbtheorem[use counter from=myTheo]{myProp}{Proposition}{
width=\linewidth,
leftrule=1.5mm,bottomrule=0mm,toprule=0mm,rightrule=0mm,arc=0mm,
breakable,toprule at break=0mm,bottomrule at break=0mm,
colback=white,colframe=black,coltitle=black,colbacktitle=white,
fonttitle=\bfseries, 
}{th}

\newtcbtheorem[]{myremark}{Remark}{
width=\linewidth,
leftrule=1.5mm,bottomrule=0mm,toprule=0mm,rightrule=0mm,arc=0mm,
breakable,toprule at break=0mm,bottomrule at break=0mm,
colback=white,colframe=black,coltitle=black,colbacktitle=white,
fonttitle=\bfseries, 
}{th}

\newtcbtheorem[]{myObs}{Observation}{
width=\linewidth,
leftrule=1.5mm,bottomrule=0mm,toprule=0mm,rightrule=0mm,arc=0mm,
breakable,toprule at break=0mm,bottomrule at break=0mm,
colback=white,colframe=black,coltitle=black,colbacktitle=white,
fonttitle=\bfseries, 
}{th}

\providecommand{\sphline}[1]{
  \noalign{\vskip #1 pt}
  \hline
  \noalign{\vskip #1 pt}
}

\title{The Stochastic Critical Node Problem over Trees}

\author
{Pierre Hosteins \thanks{Corresponding author.}\\ 
University Lille Nord de France, IFSTTAR, COSYS, ESTAS, \\
F-59650 Villeneuve d'Ascq, Lille, France \\
e-mail: pierre.hosteins@ifsttar.fr \\
\\ 
Rosario Scatamacchia\\
Politecnico di Torino, \\
Dipartimento di Ingegneria Gestionale e della Produzione, \\ 
Corso Duca degli Abruzzi 24 - 10129 Torino, Italy\\
e-mail: rosario.scatamacchia@polito.it
}

\date{\today}

\date{}

\begin{document}

\maketitle

\begin{abstract}
We tackle a stochastic version of the Critical Node Problem (CNP) where the goal is to minimize the pairwise connectivity of a graph by attacking a subset of its nodes. 
In the stochastic setting considered, the attacks on nodes can fail with a certain probability.  
In our work we focus on trees and demonstrate that over trees the stochastic CNP actually generalizes to the stochastic Critical Element Detection Problem where attacks on edges can also fail with a certain probability. We also prove the NP-completeness of the decision version of the problem when connection costs are one, while its deterministic counterpart was proved to be polynomial. We then derive linear and nonlinear models for the considered CNP version. Moreover, we propose an exact approach based on Benders decomposition and test its effectiveness on a large set of instances. As a side result, we introduce an approximation algorithm for a problem variant of interest.
\end{abstract}

\section{Introduction}

Network interdiction problems have been intensively investigated in the recent years~\cite{Smith2013}. Real-life applications of these problems arise in many fields, like e.g. transportation networks \cite{Jenelius2006537}, power distribution networks \cite{Salmeron2004,Seo2015}, diffusion phenomena such as viral infections and their mitigation \cite{Zhou2006,Ventresca2014b}, homeland security \cite{homelandsec} and bioinformatics \cite{Tomaino2012}. We refer to \cite{Smith2013} for a comprehensive introduction on network interdiction problems and related applications. 
A mainstream network interdiction problem is the Critical Element Detection Problem (CEDP) which calls for fragmenting an undirected graph as much as possible by the deletion of a subset of its edges and nodes. The aim is to give indications on the critical elements of a network to protect or to attack according to the application of interest. There exist many different connectivity metrics for estimating the dismantling of a network. 
Earlier works on network interdiction mainly focused on the maximum flow that could be transported from a source node $s$ to a sink node $t$ in a graph \cite{Wood93}, or on increasing the length of the shortest path between $s$ and $t$ \cite{Ball1989,Israeli2002}.
Studies \cite{Corley1974,Corley1982} consider further variants of max flow and shortest path problems involving the detection of the most vital nodes in a network.
 Later, the attention has focused on measures related to the cohesive properties of a graph, such as the number of maximally connected components, the maximum cardinality of the connected components or the pairwise connectivity (i.e. the number of pair of nodes still connected by a path after the deletion of a subset of nodes/edges). The Critical Node Detection Problem (CNDP) introduced in \cite{Aru-al09cnp}, also known as the Critical Node Problem (CNP), was one of the main works considering these connectivity measures. 

The Critical Node Problem focuses exclusively on node deletion and has been object of numerous publications in the recent years. 
The study \cite{Aru-al09cnp} considers the minimization of the pairwise connectivity of a graph subject to a limit on the number of nodes that can be deleted. 
The authors of \cite{Aru-al09cnp} prove the strong NP-hardness of the problem and propose a Mixed Integer Linear Programming (MILP) formulation along with a heuristic algorithm. Improved MILP formulations have been proposed afterwards in \cite{DiSumma2012, Pavlikov2018, Veremyev2014} and in \cite{Veremyev2014b} where the deletion of edges is also considered. Since the MILP formulations can only be reasonably applied to graphs of limited size, many heuristic approaches were proposed for dealing with large graphs. Among others, we mention the recent studies in \cite{INOC2015,Genetic, Networks,GRASP-CNP2, Glover2018}. 
The study of the CNP on specific graph classes such as trees attracted particular attention as well. The authors of \cite{DisGroLoc11cnp} study the complexity of the pairwise connectivity CNP over trees by analyzing different problem variants and providing polynomial and super-polynomial algorithms. The works~\cite{Lalou2016} and \cite{SmithShen12} provide complexity results for the CNP over trees and other specially structured graphs using alternative connectivity metrics, such as the maximum cardinality of the connected components, the number of connected components or the maximum pairwise connectivity between all components. Further complexity and approximation results for general or specially structured graphs are given in \cite{Addis13}. The study~\cite{Veremyev2015} discusses a generalization of the pairwise connectivity CNP where the distances between node pairs impact the objective function. The related complexity over trees is analyzed in \cite{DAM-DCNP}. More precisely, the authors of \cite{DAM-DCNP} establish complexity results according to specific distance functions and introduce polynomial and pseudo-polynomial algorithms for special graph classes such as paths, trees and series-parallel graphs. This version of the CNP also corresponds to a generalized multiple source-sink shortest path interdiction problem based on node removal and with a non-linear objective function.
Few decomposition approaches have been explored to solve the CNP through advanced Mathematical Programming methods, with the notable exceptions of~\cite{Granata2013} and~\cite{Walteros2018} which use Column Generation techniques. In these studies, the set of critical nodes is assumed to have a specific structure such as a path, a clique or a star. Benders Decomposition (BD) is used in \cite{Hooshmand2019,IJoCRobust} to respectively solve a distance-based version of the CNP and a stochastic version with uncertain costs, denoted as the robust critical node selection problem.
We refer the interested reader to the recent survey \cite{Lalou2018} for a detailed description of different CNP variants and related theoretical results and algorithms available in the literature.  

Stochastic versions of several interdiction problems were also considered in the literature.
A stochastic version of a Max-Flow Interdiction Problem is tackled in \cite{Cormican1998}. The problem calls for minimizing the maximum flow between a source and sink in a graph by removing some of its edges where each edge attack can fail with a given probability. The aim is thus to minimize the \emph{expected value} of the maximum flow over all possible scenarios of attack failures.  The authors of \cite{Cormican1998} design a sophisticated and effective method to solve the problem by partitioning the scenarios into clusters and reaching solutions with a predefined accuracy. The method 
relies on different formulations of the classical maximum flow problem with interdicted arcs with identical optimal solutions, providing respectively upper and lower bounds on the optimal solution. The algorithm allows the authors to solve the Stochastic Programming formulation without sampling over an exponential set of scenarios. However the approach exploits the specific structure of the max flow problem and can hardly be generalized to other network interdiction problems such as the CNP.

The study \cite{stoch-CNP_Thai2015} tackles a stochastic version of the CNP where the presence of the edges in a given graph is uncertain. Each edge has an independent probability to be absent from the graph. This gives $2^m$ different scenarios to analyze where $m$ denotes the number of edges in the input graph.
The goal is to minimize the expected pairwise connectivity after removing a subset of the nodes subject to a budget constraint. A MILP model is introduced with an exponentially large number of variables due to the number of possible scenarios. A heuristic approach is proposed by solving a reduced MILP with variables and constraints aggregated. 
A local search procedure is then applied to improve the computed solutions. It uses a Fully Polynomial Randomized Approximation Scheme (FPRAS) to estimate the objective value of each computed solution within a given precision in polynomial time. Finally, a CNP variant with nondeterministic connection costs for each pair of nodes was introduced in \cite{PardalosRobust,IJoCRobust} and formulated as a robust optimization problem. 

To the best of our knowledge, no solution approach exists in the literature for the stochastic version of the CNP where the attacks on nodes can fail with a certain probability. In this paper, we consider such a CNP version over trees and propose exact methods based on Mathematical Programming. We remark that the derived algorithms do not rely on any statistical approximation such as the Sample Average Approximation. In a sense, this work extends the results in the literature for the CNP over trees within a stochastic framework. From a practical point of view, the problem might find application in interdiction problems for networks with an intrinsic hierarchical structure where each node has ``superiors'' and ``subordinates'', such as the dismantling of a network of terrorists (see, e.g., \cite{DisGroLoc11cnp}). \\
The paper is organized as follows. We introduce a general Integer Linear Programming (ILP) formulation for the stochastic CNP in Section~\ref{sec:Form}. In Section~\ref{sec:CNPtree}, we present models and theoretical results for the problem variant over trees. For this variant we then propose an exact solution approach based on Benders Decomposition in Section~\ref{sec:CR_BD}. We discuss the computational performance of the proposed approach on a large set of instances in Section~\ref{sec:numerical}. We also present an ILP model and an approximation algorithm for specific problem variants in Appendix~\ref{app:ILP_equal_p} and Appendix~\ref{app:approx}. Section~\ref{sec:conclusion} provides some concluding remarks. 

\section{Notation and problem formulation over general graphs}
\label{sec:Form}

In the stochastic CNP, hereafter denoted as $SCNP$, we are given an undirected graph $G=(V,E)$ with set of nodes $V$ and set of edges $E$. Let us denote by $n := |V|$ and $m := |E|$ the number of nodes and edges respectively. Sticking to the common terminology adopted in previous studies on the CNP, we also denote by $c_{ij}$ the cost (or weight) of a connection between nodes $i$ and $j$ $(i,j\in V)$.
This parameter can be considered a cost as the presence of a connection in the CNP penalizes the corresponding objective function.
Each node $i$ ($i \in V$) has a set of neighbors $N_i := \{k: (i,k) \in E \}$, an attack cost $\kappa_i$, and probability of survival $p_i$ in a case of attack.  We have a budget value for the attacks on nodes denoted by $K$. We assume that the survival probabilities are independent. This assumption is reasonable as in many real-life applications the vulnerability of site in a network does not depend on the outcomes of attacks on sites in other parts of the network, for instance when all attacks on nodes occur simultaneously. 
The problem calls for minimizing the expected value of the pairwise connectivity in the graph after attacking a subset of the nodes. The $SCNP$ is a generalization of the deterministic CNP (which is strongly NP-hard~\cite{Addis13}) where $p_i = 0$ for all $i \in V$, i.e. an attacked node is always removed from the graph. 
We can derive an Integer Linear Programming (ILP) formulation for the problem as follows. As customary in Stochastic Programs we consider the set of possible scenarios $\S$ where each scenario defines which nodes would survive in a case of an attack. Thus the number of scenarios is exponentially large with $n$, i.e.  $|\S|=2^n$. The probability that a scenario $s \in \S$ occurs is denoted by $\pi_s$, with $\pi_s = \prod\limits_{i = 1}^n \gamma_i$ and where $\gamma_i = 1 - p_i$ if node $i$ will not survive in the scenario or else $\gamma_i = p_i$. We then introduce binary variables $v_i$ $(i=1,\dots,n)$ equal to 1 if and only if node $i$ is attacked 
and binary variables $u_{ij}^s$ equal to 1 if and only if two nodes $i$ and $j$ are connected by a path in scenario $s$ in the induced subgraph $G[V\setminus S]$, with $S :=\{i\in V: v_i=1\}$. We also introduce parameter $\delta_i^s$, which is equal to 1 if an attack on a node $i\in V$ is unsuccessful in the scenario $s\in\S$. Using for example the compact MILP formulation provided in \cite{Veremyev2014} for the deterministic CNP, the problem can be formulated as follows: 
\begin{subequations}
\label{Eq:fullmodel}
\begin{align}
\min\quad   & \sum_{s\in\S} \pi_s \sum_{i<j}c_{ij}u_{ij}^s				\label{Eq:Obj1}\\
\text{s.t.}\quad & \sum_i \kappa_i v_i \leq K                                           \label{Eq:budget1}\\
            & v_i = 0  									& i \in V: p_i = 1 \label{Eq:Pone} \\
            & u_{ij}^s \geq 1-(1-\delta_i^s)v_i-(1-\delta_j^s)v_j			& (i,j)\in E, s \in\S \label{Eq:const1} \\
            & u_{ij}^s \geq \frac{1}{|N_i|}\sum_{k\in N_i}u^s_{kj} - (1-\delta_i^s)v_i	& (i,j)\notin E: i<j, s\in\S \label{Eq:const2}\\
            & u_{ij}^s \in\{0,1\}  & i,j\in V: i<j, s \in\S \label{Eq:varUij1} \\
            & v_i\in\{0,1\} & i\in  V  \label{Eq:varVi1}
\end{align}
\end{subequations}
The objective function \eqref{Eq:Obj1} minimizes the expected value of the pairwise connectivity by going over all possible scenarios. Constraint \eqref{Eq:budget1} represents the budget constraint for the attacks on nodes in the graph. 
Constraints \eqref{Eq:Pone} enforce the fact that no node with probability of survival equal to one will be ever attacked, as it is always suboptimal to consume budget for attacking a node that would survive anyway. 
In each scenario $s$, where both nodes $i$ and $j$ of an edge $(i,j) \in E$ can be successfully removed by an attack  $(\delta_i^s = \delta_j^s = 0)$, constraints \eqref{Eq:const1} ensure that $u_{ij}^s=1$ only if both node $i$ and $j$ are not attacked $(v_i = v_j = 0)$. 
Note that when $\delta_{i}^s$ (or $\delta_{j}^s$) gets the value of one, node $i$ (or $j$) is functional in the graph even after it is attacked. Similarly, constraints \eqref{Eq:const2} guarantee that, for two nodes $i$ and $j$ not connected by an edge, we have $u_{ij}^s=1$ if there is at least one path between $j$ and a neighbor of $i$ and node $i$ remains functional in the graph (either $v_{i}=0$ and $\delta_{i}^s = 0$, or $\delta_{i}^s = 1$).
Constraints $\eqref{Eq:varUij1}$ and $\eqref{Eq:varVi1}$ define the domain of definition of the variables. Notice that the ILP formulation for the deterministic CNP where $p_i = 0$ ($i=1,\dots,n$) is equivalent to model \eqref{Eq:Obj1}-\eqref{Eq:varVi1} with one scenario $s$ with $\pi_s = 1$ and $\delta_i^s = 0$ for $i=1,\dots,n$. Notice also that model \eqref{Eq:Obj1}-\eqref{Eq:varVi1} can be applied to any graph. \\
\smallskip
However we remark that the model has a number of variables and constraints which is exponential in the number of nodes $n$ and thus it is intractable to solve even for graphs of very limited size. In small graphs with $n = 20$ the number of scenarios to handle ($2^{20} \approx 10^6$) would be already computationally prohibitive with model \eqref{Eq:Obj1}-\eqref{Eq:varVi1}. Our goal is to overcome the limits of the generic model \eqref{Eq:Obj1}-\eqref{Eq:varVi1} when the graphs considered are trees   
by offering effective exact solution approaches to deal with larger graphs and without considering any statistical approximation such as the sampling of the scenarios.

\subsection{Valid inequalities}
\label{ValIneq}

For our algorithmic developments over trees, we generalize a result for the deterministic version of the CNP where both connection costs and deletion weights are one. Several works \cite{SmithShen12,Veremyev2014,Veremyev2014b} point out that there always exists an optimal solution for such a CNP variant where no node $i\in V$ with only one neighbour ($|N_i|=1$, i.e. a \emph{leaf node}) is selected. If a solution selects a leaf node, another solution which is never worse can be obtained by replacing the leaf node with its neighbour. 

We now generalize this result to the SCNP with arbitrary connection costs $c_{ij}\geq 0$ and arbitrary attack costs $\kappa_i>0$. 
Let $D_1$ denote the set of leaf nodes, i.e. $D_1=\{i\in V: |N_i| = 1\}$. The following proposition holds.


\begin{myProp}[theorem style=plain]{}{prop:valid_ineq}
For any node pair $i\in D_1$ and $j\in N_i$, with $j \notin D_1$,  if $p_j \leq p_i$ and $\kappa_j  \leq \kappa_i$, 
then $v_i\leq v_j$ in at least one optimal solution of model \eqref{Eq:Obj1}-\eqref{Eq:varVi1}.
\end{myProp}
\begin{proof}
We consider the value of $v_i$ and $v_j$ in an optimal solution. If nodes $i$ and $j$ are attacked, i.e. $v_i = v_j = 1$, or both nodes are not attacked, i.e. $v_i = v_j = 0$, then we have $v_i=v_j$ and the claim holds.   
So we now compare the two remaining cases with $v_i\neq v_j$ where only one node between $i$ and $j$ is attacked.  \\
Consider first the case where only node $i$ is attacked, i.e. $v_i = 1$ and $v_j=0$. 
Let us call $\langle u^{(1)}_{kl}\rangle$ the average probability that nodes $k$ and $l$ are connected by a path in this configuration, namely $\langle u^{(1)}_{kl}\rangle =\sum_{s\in\S}\pi_su^s_{kl}$. 
The part of the objective function involving nodes $i$ and $j$ is given by $\sum_{k\in V:k\neq i,j}c_{kj}\langle u^{(1)}_{kj}\rangle+\sum_{k\in V:k\neq i}c_{ki}\langle u^{(1)}_{ki}\rangle$. Since any path between $i$ and another node $k$ has to go through $j$, 
we have that $\langle u^{(1)}_{ki}\rangle=p_i\langle u^{(1)}_{kj}\rangle$ for any $k\neq i,j$. Using the fact that $\langle u^{(1)}_{ij}\rangle=p_i$, we have: 
\begin{equation}
\label{TheFirstSol}
\sum_{k\in V:k\neq i,j}c_{kj}\langle u^{(1)}_{kj}\rangle+\sum_{k\in V:k\neq i}c_{ki}\langle u^{(1)}_{ki}\rangle = p_ic_{ij}+\sum_{k\in V:k\neq i,j}(c_{kj}+p_ic_{ki})\langle u^{(1)}_{kj}\rangle.
\end{equation}
Consider now the case where only node $j$ is attacked, $v_i=0$ and $v_j=1$, and denote by $\langle u^{(2)}_{kl}\rangle$
the average probability that two nodes $k$ and $l$ are connected by a path. We have $\langle u^{(2)}_{ki}\rangle=\langle u^{(2)}_{kj}\rangle$ for $k\neq i,j$ and $\langle u^{(2)}_{ij}\rangle=p_j$. 
By also considering that $\langle u^{(2)}_{kj}\rangle=p_j\langle u^{(1)}_{kj}\rangle$ for $k\neq i,j$, the contribution in the objective function associated with nodes $i$ and $j$ is:
\begin{equation}
\label{TheSecSol}
\sum_{k\in V:k\neq i,j}c_{kj}\langle u^{(2)}_{kj}\rangle+\sum_{k\in V:k\neq i}c_{ki}\langle u^{(2)}_{ki}\rangle = p_jc_{ij}+ \sum_{k\in V:k\neq i,j}(p_jc_{kj}+p_jc_{ki})\langle u^{(1)}_{kj}\rangle.
\end{equation}
Notice that the values of all the other terms associated with the remaining nodes in the objective function do not change when only the values of $v_i$ and $v_j$ change. Thus the difference in the objective value provided by the two solutions corresponds to the difference between \eqref{TheFirstSol} and \eqref{TheSecSol}, i.e.:
\begin{equation}
\label{DiffSol}
 (p_i - p_j)c_{ij}+\sum_{k\in V:k\neq i,j}((1 - p_j)c_{kj}+(p_i-p_j)c_{ki})\langle u^{(1)}_{kj}\rangle
\end{equation}
which is nonnegative when $p_j \leq p_i$ as terms $c_{ij}$, $c_{kj}$, $\langle u^{(1)}_{kj}\rangle$ are nonnegative. Hence, since $\kappa_j \leq \kappa_i$, a solution that sets $v_i = 1$ and $v_j = 0$ can be improved by attacking node $j$ instead of $i$, i.e. $v_i = 0$ and $v_j = 1$, which implies $v_i \leq v_j$.
\end{proof}

Although the rest of this work is devoted to the study of trees, we stress that the previous inequalities are valid for general graphs.

\section{CNP with stochastic node removal on trees}
\label{sec:CNPtree}

We now turn our attention to the stochastic CNP over trees where the outcome of a node attack is uncertain. We denote this problem as \Trees. We also denote as $D$-\Trees  the decision version of the problem asking whether there exists a solution of the \Trees with a value not superior to a target value $\Gamma$. In this section we first present a nonlinear reformulation of the problem which constitutes the basis for deriving a Mixed Integer Linear Programming (MILP) model and an exact approach introduced in Section~\ref{sec:CR_BD}. We also prove the NP-completeness of the $D$-\Trees even when connection costs are one, while its deterministic counterpart was proved to be polynomial. We finally discuss some generalizations to other stochastic problems over trees.

\subsection{Nonlinear reformulation}
Two given nodes in a tree are connected by a unique path of nodes. Hence, the expected cost of the connection between two nodes $i$ and $j$ only depends on the products between $c_{ij}$ and the survival probabilities of the attacked nodes in their path.
We denote by $\P_{ij}$ the set of nodes, including nodes $i$ and $j$, in the path between $i$ and $j$. We also denote by $S_{ij}$ the set of nodes which are attacked in $\P_{ij}$, i.e. $S_{ij}=S\cap\P_{ij}$. Since the expected cost of the connection between $i$ and $j$ is equal either to $c_{ij}$ if $S_{ij}$ is empty or to the product $c_{ij}\prod_{k\in S_{ij}}p_k$, we can state a nonlinear model 
for \Trees with a polynomial number of variables and constraints. By keeping the notation of model \eqref{Eq:Obj1}-\eqref{Eq:varVi1}, we consider only binary variables $v_i$ associated with the attack of a node $i$ and introduce the following nonlinear reformulation:
\begin{subequations}
\label{Eq:Stoch-INLP}
\begin{align}
\min\quad   & \sum_{i<j}c_{ij}\prod_{k\in \P_{ij}}(1-(1-p_k)v_k)	& \label{Eq:Obj3}\\
\text{s.t.}\quad & \sum_i \kappa_iv_i \leq K				& \label{Eq:budget3}\\
            & v_i = 0							& i \in V: p_i = 1 \label{Eq:Pone3} \\
            & v_i \leq v_j						& i \in D_1, \; j\in N_i, \; j \notin D_1, \; p_j \leq p_i, \; \kappa_j\leq\kappa_i \label{Eq:ValidIneq}\\
            & v_i\in\{0,1\}						& i\in  V. \label{Eq:Var3}
\end{align}
\end{subequations}
The objective function \eqref{Eq:Obj3} represents the same sum over the connection costs as in objective \eqref{Eq:Obj1}. However, we have already performed the sum over the exponential set of scenarios $\S$. Since only path $\P_{ij}$ exists between two given nodes $i$ and $j$, the probability of survival of the connection corresponds to the probability of survival of the path. Such a probability can be computed by multiplying the probabilities of survival of the nodes, given that these probabilities are independent. In particular, for each node $k \in \P_{ij}$ the probability of survival is either 1 if the node is not attacked ($v_k=0$) or $p_k$ if the node is attacked ($v_k=1$). Therefore the product $\prod_{k\in \P_{ij}}(1-(1-p_k)v_k)$ in \eqref{Eq:Obj3} represents the product $\prod_{k\in S_{ij}}p_k$. 
Constraints~\eqref{Eq:budget3} and \eqref{Eq:Pone3} are the same as constraints~\eqref{Eq:budget1} and \eqref{Eq:Pone}, respectively. Finally, constraints \eqref{Eq:ValidIneq} represent the valid inequalities introduced in Section~\ref{ValIneq}. 
As discussed in detail in Section~\ref{sec:CR_BD}, we propose a linear reformulation of model \eqref{Eq:Obj3}-\eqref{Eq:Var3}, that allows us to develop an effective exact approach based on Benders Decomposition. We also manage to derive a linear formulation for the specific problem variant where all survival probabilities are equal, i.e $p_i = p$ for $i\in V$. We present the corresponding ILP model in Appendix~\ref{app:ILP_equal_p}.

\subsection{NP-completeness with unit connection costs}
We now discuss the complexity of the $D$-\Trees. The expected cost of the connection between two nodes $i$ and $j$ can be computed in $O(n)$ as it only depends on the products between $c_{ij}$ and the survival probabilities of the attacked nodes in their path. By considering all $O(n^2)$ paths in the tree, the overall objective value of each given solution can be computed in polynomial running time $O(n^3)$, implying that the $D$-\Trees is an NP problem. We manage to prove the NP-completeness of the $D$-\Trees even with unit connection costs, i.e. $c_{ij} = 1$ for all $i,j \in V$, while the deterministic pairwise CNP variant with unit connection costs was proved to be polynomial over trees in~\cite{DisGroLoc11cnp}.

\begin{myProp}[theorem style=plain]{}{prop:edge_uncertainty}
The $D$-\Trees with unit connection costs is NP-complete.
\end{myProp}
\begin{proof}
We prove the theorem by a reduction from the Knapsack Problem ($KP$), a well-known optimization problem where a capacity value $C$ and a set of $n$ items with profits $\pi_i > 0$ and weights $w_i > 0$ are given. The goal is to select a subset of the items to maximize the profits while ensuring that the weight of the selected items does not exceed $C$. The decision version of $KP$, denoted by $D$-$KP$, is NP-complete and asks whether there exists a feasible solution with a profit not inferior to a value $\Pi > 0$.
We can map each $D$-$KP$ instance into a $D$-\Trees instance as follows. We consider a root node with survival probability $p_1=0$ and deletion cost $\kappa_1=1$. At the next level of the tree, we introduce $n$ intermediate nodes $2, \dots, (n+1)$ with survival probabilities $p_{i+1}=1$ and deletion costs $\kappa_{i+1}=1$ for $i=1,\dots, n$. Then, to each of the intermediate nodes we attach a leaf node which corresponds to an item in the knapsack instance. Each leaf node $(i+n+1)$, for $i=1,\dots,n$, has a deletion cost $\kappa_{i+n+1}=w_i$ and a survival probability $p_{i+n+1}=1-\pi_i/\pi_{max}$, with $\pi_{max} = \max_{i} \{ \pi_i \}$. Finally, we set $K = C+1$, $\Gamma = n - \frac{\Pi}{\pi_{max}}$ and $c_{ij} = 1$ for each pair of nodes $i,j$. Such a reduction is polynomial in $n$.\\
To prove the NP-completeness of the $D$-\Trees, we have to show that a $D$-$KP$ instance is a ‘‘Yes’’ instance, i.e. it has
a feasible solution with a profit not inferior to $\Pi$, if and only if the related $D$-\Trees instance is a ‘‘Yes’’ instance, i.e. the instance admits a solution with a value not superior to $\Gamma$.
Notice that in the considered settings any solution of a ‘‘Yes’’ instance of the $D$-\Trees must attack (and successfully delete) the root note. Else, a solution value of at least $\frac{(n+1)((n+1)-1)}{2} \geq n > \Gamma$ would be induced by the connections involving the root node and the intermediate nodes, which cannot be removed after an attack as their survival probability is equal to 1. Since attacks on intermediate nodes only induce budget consumptions without affecting the objective value and hence are suboptimal, without loss of generality we consider solutions of the $D$-\Trees where the intermediate nodes are not attacked. 
Let us consider a ‘‘Yes’’ instance of $D$-\Trees and a solution that attacks the root node and a subset of the leaves with the residual budget $K -1 = C$. If we denote by $I$ the index set of the attacked leaf nodes $(i+n+1)$ with $i \in I$, we get a solution with weight $1 + \sum_{i\in I} \kappa_{i+n+1} \leq K$ which implies $\sum_{i\in I} w_{i} \leq C$ and objective value 
$$\sum_{i \not\in I} 1 + \sum_{i\in I} p_{i+n+1} = \sum_{i \not\in I} 1 + \sum_{i \in I} 1  - \frac{\sum_{i\in I} \pi_i}{\pi_{max}} = n - \frac{\sum_{i\in I} \pi_i}{\pi_{max}} \leq \Gamma$$ which implies $\sum_{i\in I} \pi_i \geq \Pi$. Hence, set $I$ provides also a solution of the $D$-$KP$ instance.\\
Likewise, consider a ‘‘Yes’’  instance of $D$-$KP$ and a solution with item set $I'$, $\sum_{i\in I'} w_i \leq C$ and  $\sum_{i\in I'} \pi_i \geq \Pi$. We derive a solution of the $D$-\Trees instance by selecting the root node and leaf nodes $(i+n+1)$ with $i \in I'$ for an attack. The corresponding weight and profit entries are $1 + \sum_{i\in I'} \kappa_{i+n+1} \leq C + 1 = K$ and $n - \frac{\sum_{i\in I'} \pi_i}{\pi_{max}} \leq n - \frac{\Pi}{\pi_{max}} = \Gamma$.
\end{proof}
The inherent difficulty of solving the stochastic CNP over trees even with unit costs also motivates the development of the exact algorithm introduced in Section~\ref{sec:CR_BD}.

\subsection{Generalizations to other stochastic interdiction problems over trees}

We conclude this section by presenting a series of further results for the problem. We first show that the considered CNP variant over trees actually generalizes to the CEDP where also attacks on edges can succeed with a given probability only. The following proposition holds.

\begin{myProp}[theorem style=plain]{}{prop1}
The stochastic CEDP over trees with edge and node removal reduces to the \Trees. 
\end{myProp}
\begin{proof}
A stochastic CEDP instance has a survival probability $p_{ij}$ and a deletion cost $\kappa_{ij}$ for each edge $(i, j) \in E$, in addition to the other inputs of a \Trees instance. To obtain a \Trees instance, it suffices to associate each edge $(i,j)\in E$ with a new node $v_{ij}$ emanating two edges towards node $i$ and $j$ respectively, giving a new tree with $2n-1$ nodes. For each new node $v_{ij}$, setting $p_{v_{ij}}=p_{ij}$ and $\kappa_{v_{ij}} = \kappa_{ij}$ completes the reduction.
\end{proof}

An alternative stochastic version of the problem also comes to mind where each edge is present in the graph with an independent probability $p_{ij}$, as proposed in \cite{stoch-CNP_Thai2015}. The problem calls for the minimization of the expected value of the pairwise connectivity over all the possible realizations of the graph. However, the following proposition shows that this CNP variant might be of limited interest over trees as it can be reduced to the deterministic CNP and solved with existing methods available in the literature.
\begin{myProp}[theorem style=plain]{}{prop1}
The stochastic CNP over trees with uncertainty on the existence of edges reduces to the deterministic CNP.
\end{myProp}
\begin{proof}
In a tree the probability that a connection exists between two nodes $i$ and $j$ is equal to $\prod_{(k,l)\in \E_{ij}}p_{kl}$ where $\mathcal{E}_{ij}$ denotes the set of edges in the unique path between $i$  and $j$.  It is sufficient to redefine the connection costs $c_{ij}$ as $c^\prime_{ij} = c_{ij}\prod_{(k,l)\in \E_{ij}}p_{kl}$ so as to obtain an instance of the deterministic CNP. 
\end{proof}

\section{An exact solution approach for the \Trees}
\label{sec:CR_BD}

We propose an exact solution approach based on a linearization of model \eqref{Eq:Obj3}-\eqref{Eq:Var3} and on BD. We first present a MILP reformulation of model \eqref{Eq:Obj3}-\eqref{Eq:Var3} and then introduce the proposed approach.

\subsection{MILP reformulation of model \eqref{Eq:Obj3}-\eqref{Eq:Var3}}
We can linearize the objective function of model \eqref{Eq:Obj3}-\eqref{Eq:Var3} by introducing additional variables and constraints in order to compute the probability that a given connection exists through a recursive analysis of the corresponding attacked nodes. More precisely, we employ the so-called \emph{probability chains} from probability theory (see, e.g., \cite{OHanley2013}).  
We compute the surviving probability of each connection $i$-$j$ as follows.
Let us introduce continuous variables $s^{ij}_k$, with $0 \leq s^{ij}_k \leq 1$ and $k\in  \P_{ij}$, representing the probability that a connection $i$-$j$ still exists when possible attacks on the first $k$ nodes in $\P_{ij}$ are considered. We also introduce continuous variables $r^{ij}_k$, with $0 \leq r^{ij}_k \leq 1$ that represent the probability of disconnecting nodes $i$ and $j$ by considering an attack on the $k-th$ node in the path, provided that the connection survived to possible attacks on the previous nodes. From now on, we will consider the nodes in a path $i$-$j$ indexed in increasing order. Correspondingly, we indicate by $v_{[1]}, v_{[2]}, \dots,  v_{[|\P_{ij}|]}$ the binary variables associated with attacks on the nodes in the original tree and by $p_{[1]}, p_{[2]}, \dots,  p_{[|\P_{ij}|]}$ the related probabilities of survival. For the first node in $\P_{ij}$ we have:

$$r^{ij}_1 = (1-p_{[1]})v_{[1]},$$
$$s^{ij}_1  =  1 - r^{ij}_1.$$

Moving to the second node in the path, by definition of variables $r$, we have that 
$$r^{ij}_2 = (1 - p_{[2]})v_{[2]}s^{ij}_1.$$
Since variables $v$ are binary and variables $r$ and $s$ are between 0 and 1, the previous nonlinear expression can be linearized by the following set of constraints:
\begin{subequations}
\begin{align}
r^{ij}_2 \leq (1 - p_{[2]})v_{[2]}, \nonumber \\
r^{ij}_2 \leq (1 - p_{[2]})s^{ij}_1,  \nonumber \\
r^{ij}_2 \geq (1 - p_{[2]})(s^{ij}_1 + v_{[2]} -1). \nonumber
\end{align}
\end{subequations}
The probability of survival $s^{ij}_2$ is then equal to $1$ minus the probability that at least a previous attack is successful, i.e.:
\begin{subequations}
\begin{align}
s^{ij}_2 = 1 - ((1 - s^{ij}_1)+ r^{ij}_2) = s^{ij}_1 - r^{ij}_2. \nonumber
\end{align}
\end{subequations}

By applying the same recursive argument for all nodes in the path, we compute the overall probability $s^{ij}_{|\P_{ij}|}$ that connection $i$-$j$ exists, i.e. $s^{ij}_{|\P_{ij}|}$ = $\prod_{k\in \P_{ij}}(1-(1-p_k)v_k)$. Hence, we linearize the objective function of model  \eqref{Eq:Obj3}-\eqref{Eq:Var3} by replacing the nonlinear terms in the objective with $s^{ij}_{|\P_{ij}|}$ and by adding the corresponding constraints and variables for each path $i$-$j$. We obtain the following MILP model:

\begin{subequations}
\begin{align}
\min\quad & \sum_{i<j}c_{ij}s^{ij}_{|\P_{ij}|}			& \label{Eq:THEObj}\\
\text{s.t.}\quad & \sum_i \kappa_iv_i \leq K			& \label{Eq:THEbudget}\\
& v_i = 0							& i \in V: p_i = 1 \label{Eq:THEPone} \\
& v_i \leq v_j							& i \in D_1, \; j\in N_i, \; j \notin D_1, \; p_j \leq p_i, \; \kappa_j\leq\kappa_i \label{Eq:THEValidIneq}\\
& r^{ij}_1 = (1-p_{[1]})v_{[1]}   				& i,j\in V: i<j \label{Eq:THEfirstrij}\\
& s^{ij}_1  =  1 - r^{ij}_1					& i,j\in V: i<j\\
& r^{ij}_k \leq (1 - p_{[k]})v_{[k]}				& k=2,\dots,|\P_{ij}|,\; i,j\in V: i<j \\
& r^{ij}_k \leq (1 - p_{[k]})s^{ij}_{(k-1)}			& k=2,\dots,|\P_{ij}|,\; i,j\in V: i<j \\
& r^{ij}_k \geq (1 - p_{[k]})(s^{ij}_{(k-1)} + v_{[k]} -1)	& k=2,\dots,|\P_{ij}|,\; i,j\in V: i<j \\ 
& s^{ij}_k = s^{ij}_{(k-1)} - r^{ij}_k				& k=2,\dots,|\P_{ij}|,\; i,j\in V: i<j \label{Eq:THElastsij}\\
& r^{ij}_k \geq 0, s^{ij}_k \geq 0				& k=1,\dots,|\P_{ij}|,\; i,j\in V: i<j \\
& v_i\in\{0,1\} &\qquad i\in  V	\label{Eq:THEVarV}
\end{align}
\end{subequations}
Notice that, since $0 \leq p_i \leq 1$ $(i=1, \dots, n)$, we can avoid to add bounds $r^{ij}_k \leq 1$, $s^{ij}_k \leq 1$ $(k=1,\dots |\P_{ij}|)$. Also notice that for each subpath $i'$-$j'$ of path $i$-$j$ with $i' = i$ and $j' < j$, the related constraints are already taken into account in the constraints for path $i$-$j$. Thus for these subpaths we just have to replace  term $\prod_{k\in \P_{i'j'}}(1-(1-p_k)v_k)$ with variable $s^{i'j'}_{|\P_{i'j'}|}$ without adding additional constraints.

\subsection{Benders Decomposition approach}
In model \eqref{Eq:THEObj}-\eqref{Eq:THEVarV}, any feasible assignment of values to binary variables $v_i$ induces a linear programming model related to all paths in the tree and with variables $r^{ij}$ and $s^{ij}$ only. The optimal solution of such an LP model coincides with the value of a \Trees feasible solution given by attacking nodes $i$ with $v_i = 1$, i.e. it is equal to the sum of the average costs of the connections. A crucial observation here is that an optimal solution of the induced LP model can be determined by separately considering each subproblem associated with a path between two nodes $i$ and $j$. All the subproblems for pairs $(i,j)$ are in fact independent once the variables $v_i$ are set. 
These structural aspects motivate us to apply the Benders Decomposition method \cite{Benders62} to efficiently solve model \eqref{Eq:THEObj}-\eqref{Eq:THEVarV}. The classical version of the Benders algorithm relies on the construction of an integer Master Problem (MP) involving all the binary/integer variables of a MILP model, denoted as ``complicating'' variables, and on the construction of a continuous linear Slave Problem (SP) induced by fixing the integer variables. The MP is iteratively solved providing nondecreasing lower bounds (for a minimization problem) on the optimal solution value of the original model. In each iteration an optimal solution of the MP induces a SP that can provide either a ``feasibility'' cut for the MP if SP is infeasible or else an optimality cut and a feasible solution (an upper bound) for the original problem. This procedure is repeated until an optimal solution is certified by the lower bounds provided by the MP. 
We describe the details of the proposed approach in the next subsections. We refer the interested reader to the recent survey \cite{RaCrGeRe17} and the references therein for a discussion about applications and algorithmic variants of the BD. We also mention the recent works~\cite{Hooshmand2019,IJoCRobust} for BD applications to variants of the CNP.

\subsubsection{Master Problem}
For the \Trees we construct the following Master Problem. Since for any given assignment of variables $v_i$ we can separately analyze the paths in the tree, for each path $i$-$j$ we introduce a nonnegative variable $z_{ij}$ that relates to the cost of connection $i$-$j$ through a set of constraints $\Phi_{ij}(z_{ij}, v)$. As explained next, each set $\Phi_{ij}(z_{ij}, v)$ is iteratively filled with optimality cuts induced by the slave subproblem $SP_{ij}$ associated with path $i$-$j$. 
We thus have the following formulation: 
\begin{subequations}
\label{Eq:BenderMP}
\begin{align}
\text{MP: } &					& \nonumber \\ 
 \min\quad  & \sum_{i<j}z_{ij}			& \label{Eq:ObjB}\\
\text{s.t.}\quad & \sum_i \kappa_iv_i \leq K	& \label{Eq:budgetB}\\
& v_i = 0  & i \in V: p_i = 1			& \label{Eq:PoneB} \\
& v_i \leq v_j 					& i \in D_1, \; j\in N_i, \; j \notin D_1, \; p_j \leq p_i, \; \kappa_j\leq\kappa_i \label{Eq:ValidIneqB}\\
& \Phi_{ij}(z_{ij}, v)			& i,j\in V: i<j \label{Eq:cutsB}\\
& v_i\in\{0,1\}					& i\in  V \label{Eq:VarVB}\\
& z_{ij} \geq 0					& i,j\in V: i<j \label{Eq:VarZB}
\end{align}
\end{subequations}
Solving to optimality the MP provides at each iteration a lower bound on the \Trees optimal value and an assignment of the $v_i$ variables to the slave subproblems. The values of variables $v_i$ also provide a feasible solution for the \Trees.

\subsubsection{Slave Problems}
Each Slave Problem $SP_{ij}(v)$ associated with path $i$-$j$ receives in input an assignment of the $v_i$ variables provided by the MP at each iteration and encoded in vector $v$. A problem $SP_{ij}(v)$ has term $c_{ij}s^{ij}_{|\P_{ij}|}$ in the objective function and constraints \eqref{Eq:THEfirstrij}-\eqref{Eq:THElastsij} where the terms involving variables $v_i$ become constant and are placed in the right-hand sides. We get the following LP formulation:
\begin{subequations}
\label{Eq:BenderSlaveij}
\begin{align}
\text{$SP_{ij}(v)$: }								& \nonumber \\ 
\min\quad &  c_{ij}s^{ij}_{|\P_{ij}|} \label{FOslave} \\ 
\text{s.t.}\quad &r^{ij}_1 = (1-p_{[1]} )v_{[1]}					& (\lambda^{ij}_{11}) \label{slave1} \\
&s^{ij}_1 + r^{ij}_1   =  1							& (\lambda^{ij}_{12}) \label{slave2}  \\
&r^{ij}_k \leq (1 - p_{[k]})v_{[k]}						& k=2,\dots,|\P_{ij}|\quad (\lambda^{ij}_{k1}) \label{slave3} \\
&r^{ij}_k - (1 - p_{[k]})s^{ij}_{k-1} \leq 0					& k=2,\dots,|\P_{ij}|\quad (\lambda^{ij}_{k3}) \label{slave4} \\
&r^{ij}_k - (1 - p_{[k]})s^{ij}_{k-1} \geq (1 - p_{[k]})(v_{[k]} - 1)		& k=2,\dots,|\P_{ij}|\quad (\lambda^{ij}_{k4}) \label{slave5} \\
&s^{ij}_k - s^{ij}_{k-1} + r^{ij}_k = 0						& k=2,\dots,|\P_{ij}|\quad (\lambda^{ij}_{k2}) \label{slave6} \\
&r^{ij}_k \geq 0, \; s^{ij}_k \geq 0						& k=1,\dots,|\P_{ij}| \label{defrksk}
\end{align}
\end{subequations}
In problem $SP_{ij}(v)$ we associate dual variables $\lambda^{ij}_{k\ell}$ with constraints \eqref{slave1}-\eqref{slave6}, with index $\ell = 1,2$ for $k=1$ and $\ell =1,\dots,4$ for $k > 1$. Notice that we assign the index value $l=2$ to constraints~\eqref{slave6} 
as they correspond to constraint~\eqref{slave2} for $k>1$ and in order to write the dual problem in a more compact form. 
The corresponding dual formulation of the subproblem $SP_{ij}(v)$, denoted as $SP^{dual}_{ij}(v)$, is as follows:
 \begin{subequations}
\label{Eq:BenderSlaveij}
\begin{align}
\text{$SP^{dual}_{ij}(v)$: }		& \nonumber \\ 
\max\quad & (1 - p_{[1]})v_{[1]}\lambda^{ij}_{11} + \lambda^{ij}_{12} + \sum_{k=2}^{|\P_{ij}|} ((1-p_{[k]})v_{[k]}\lambda^{ij}_{k1} + (1-p_{[k]})(v_{[k]}-1)\lambda^{ij}_{k4})  \label{DualFOslave} \\ 
\text{s.t.}\quad &\lambda^{ij}_{11} + \lambda^{ij}_{12}  \leq 0   & \label{Dualslave1} \\
&\lambda^{ij}_{k1} + \lambda^{ij}_{k3} +\lambda^{ij}_{k4} + \lambda^{ij}_{k2} \leq 0 &\quad k=2,\dots |\P_{ij}| \label{Dualslave2} \\
&\lambda^{ij}_{k2} -(1-p_{[k+1]})(\lambda^{ij}_{(k+1)3} + \lambda^{ij}_{(k+1)4}) -\lambda^{ij}_{(k+1)2} \leq 0  & \quad k=1,\dots |\P_{ij}| - 1 \label{Dualslave3} \\
&\lambda^{ij}_{|\P_{ij}|2} \leq c_{ij}  & \label{Dualslave4} \\
&\lambda^{ij}_{11}\in\mathbb{R}, \lambda^{ij}_{12}\in\mathbb{R}, \lambda^{ij}_{k2}\in\mathbb{R}:\,k=1,\dots,|\P_{ij}| & \label{Dualdefk1}\\
&\lambda^{ij}_{k1}  \leq 0, \lambda^{ij}_{k3} \leq 0, \lambda^{ij}_{k4}  \geq 0:\, k=2,\dots,|\P_{ij}| &\label{Dualdefk2}
\end{align}
\end{subequations}
Dual constraint \eqref{Dualslave1} is associated with variable $r^{ij}_1$, constraints \eqref{Dualslave2} are associated with variables $r^{ij}_k$ $(k=2, \dots,  |\P_{ij}|)$, constraints \eqref{Dualslave3} with variables $s^{ij}_k$ $(k=1,\dots,|\P_{ij}|-1)$ and constraint \eqref{Dualslave4} with variable $s^{ij}_{|\P_{ij}|}$, respectively. \\

\smallskip
The optimal solution value of problem $SP_{ij}(v)$ (or of its dual problem $SP^{dual}_{ij}(v)$) yields the cost of the connection $i$-$j$ for a feasible assignment of variables $v_i$ provided by the MP. We remark that, given the values of variables $v_i$, $SP_{ij}(v)$ is a feasible problem admitting the following straightforward optimal solution
\begin{align}
&s^{ij}_{k} =\prod_{\ell \leq k}(1-(1-p_{[\ell]})v_{[\ell]}) & k=1,\dots, |\P_{ij}| \label{OptSPijS}\\
&r^{ij}_k = (1 - p_{[k]})v_{[k]}s^{ij}_{k-1}  = (1 - p_{[k]})v_{[k]}\prod_{\ell \leq k-1}(1-(1-p_{[\ell]})v_{[\ell]}) &  k=1,\dots,|\P_{ij}| \label{OptSPijR}
\end{align}
with objective value $c_{ij}s^{ij}_{|\P_{ij}|} =c_{ij} \prod_{\ell \leq |\P_{ij}|}(1-(1-p_{[\ell]})v_{[\ell]})$. Notice that the last product in \eqref{OptSPijR} is equal to 1 when $k=1$. No feasibility cut for the MP is hence derived from $SP_{ij}$. Let us now denote by $\hat{\lambda}^{ij}$ the value of variable $\lambda^{ij}$ in an optimal solution of the dual problem $SP^{dual}_{ij}(v)$. If the corresponding solution value in  \eqref{DualFOslave} ($= c_{ij}s^{ij}_{|\P_{ij}|}$ by strong duality) is superior to the value of variable $z_{ij}$ in the MP, we add the following optimality cut to set $\Phi_{ij}(z_{ij}, v)$ in the MP:
\begin{align}
z_{ij} \geq (1 - p_{[1]})v_{[1]}\hat{\lambda}^{ij}_{11} + \hat{\lambda}^{ij}_{12} + \sum_{k=2}^{|\P_{ij}|} ((1-p_{[k]})v_{[k]}\hat{\lambda}^{ij}_{k1} + (1-p_{[k]})(v_{[k]}-1)\hat{\lambda}^{ij}_{k4}), \label{theOptcut}
\end{align}
as such a constraint violates the current optimal solution of the MP. We consider all the slave subproblems at each iteration and possibly add a cut for each subproblem. This strategy is often indicated in the literature as a multi-cut reformulation (see e.g. \cite{RaCrGeRe17}). Then the MP is solved again and the process repeats until the lower bound converges to the value of the best \Trees solution found. The pseudo code of the proposed exact approach is sketched below. After an initialization step (Lines \ref{init}-\ref{emptyphi}) where we start with empty sets $\Phi_{ij}(z_{ij}, v)$, we iteratively solve the MP (Line \ref{solveMP}) as long as it provides a lower bound $LB$ inferior to the best solution value $UB$ obtained so far within a given precision value $\varepsilon$ (\textit{while-loop} in Lines \ref{whileLoop}-\ref{endWhile}). At each iteration the value of LB and UB are updated according to the solutions given by the MP (Lines \ref{UpdateLB}-\ref{UpdateUB}). We also add the relevant cuts to the MP given by solving the dual slave subproblems (Lines \ref{forcuts}-\ref{endforcuts}). The solution set $S$ of the attacked nodes in the best solution is finally returned (Line \ref{returnS}). 
\begin{algorithm}[H]
\setstretch{1.13}
\label{algo:ExAp}
\caption{Algorithm \BDAL}
	\begin{algorithmic}[1]
	\State{\textbf{Input:} \Trees instance, parameter $\varepsilon$;}
	\State{$S:=\{ \emptyset \}$; $LB:=0$; $UB:=\infty$;  \label{init}}
	 \State{$\Phi_{ij}(z_{ij}, v) :=\{ \emptyset \}$ for $i,j\in V$: $i<j$; \label{emptyphi}}
	
 \While{$UB-LB>\varepsilon$} \label{whileLoop}
 \State{$(\tilde{z},\tilde{v}) \leftarrow$ solve MP; \label{solveMP}}
   \IIf{$\sum_{i<j}\tilde{z}_{ij} > LB$} $LB := \sum_{i<j}\tilde{z}_{ij}$; \EndIIf \label{UpdateLB}
    \For{$i,j\in V$: $i<j$} 
    \State{$(s^{ij}, r^{ij}) \leftarrow$ compute an optimal $SP_{ij}(\tilde{v})$ solution according to \eqref{OptSPijS}-\eqref{OptSPijR};} \EndFor
    \IIf{$\sum_{i<j} c_{ij}s^{ij}_{|\P_{ij}|} < UB$}
     $UB := \sum_{i<j} c_{ij}s^{ij}_{|\P_{ij}|}$,\;
     $S :=\{i\in V: \tilde{v}_i=1\}$;\;
   \EndIIf  \label{UpdateUB}
   \For{$i,j\in V$: $i<j$} \label{forcuts}
     \If{$\tilde{z}_{ij}< c_{ij}s^{ij}_{|\P_{ij}|}$}
      \State{ $\hat{\lambda}^{ij} \leftarrow$ solve $SP^{dual}_{ij}(\tilde{v})$; \label{SolveTheDualSP}}
       \State{Add constraint \eqref{theOptcut} to set $\Phi_{ij}(z_{ij}, v)$; \label{AddthecutSP}}
   \EndIf
        \EndFor \label{endforcuts}
 \EndWhile \label{endWhile}
 \State{\Return $S$. \label{returnS}}
\end{algorithmic}
\end{algorithm}


\subsubsection{Analytical derivation of an optimal solution of the dual slave problems}
We can analytically compute the optimal solution values $\hat{\lambda}^{ij}$ (Line \ref{SolveTheDualSP} in the pseudo code of algorithm \BDAL) for each slave subproblem  $SP^{dual}_{ij}(v)$ by the following procedure denoted as $ComputeSol\lambda^{ij}$. The derived procedure allows us to quickly derive cuts for the MP and to speed up the convergence process of the algorithm.

\begin{algorithm}[H]
\setstretch{1.13}
 \label{algo:dual_var}
\caption{Procedure $ComputeSol\lambda^{ij}$}
	\begin{algorithmic}[1]
	\State{\textbf{Input:} A dual slave problem $SP^{dual}_{ij}(v)$ corresponding to an MP solution vector $v$;}
  \State{Set $\hat{\lambda}^{ij}_{1\ell} = 0$ for $\ell=1,2$ and $\hat{\lambda}^{ij}_{k\ell} = 0$ for $k=2,\dots,|\P_{ij}|$ and $\ell=1,2,3,4$;}
   \For{$i=1$ to $i=n$}
  \IIf{($v_i = 1$ and $p_i = 0$)} \Return $\hat{\lambda}^{ij}_{k\ell}$ for $k=1,\dots,|\P_{ij}|$ and $\ell=1,2,3,4$;  \EndIIf
 \EndFor
  \State{$\hat{\lambda}^{ij}_{|\P_{ij}|1} = c_{ij}(v_{[|\P_{ij}|]}-1)$;}
  \State{$\hat{\lambda}^{ij}_{|\P_{ij}|2} = c_{ij}$;}
  \State{$\hat{\lambda}^{ij}_{|\P_{ij}|3} = -c_{ij}v_{[|\P_{ij}|]}$;}
  \State{$\hat{\lambda}^{ij}_{|\P_{ij}|4} = 0$;}
   \For{$k=|\P_{ij}| - 1$ to $k=2$}
    \State{$\hat{\lambda}^{ij}_{k2} = (1-p_{[k+1]})(\hat{\lambda}^{ij}_{(k+1)3} + \hat{\lambda}^{ij}_{(k+1)4}) +\hat{\lambda}^{ij}_{(k+1)2}$;}
   \State{$\hat{\lambda}^{ij}_{k1} = (v_{[k]}-1)\hat{\lambda}^{ij}_{k2}$;}
      \State{$\hat{\lambda}^{ij}_{k3} = -v_{[k]}\hat{\lambda}^{ij}_{k2}$;}
    \State{$\hat{\lambda}^{ij}_{k4} = 0$;}
    \EndFor
   \State{$\hat{\lambda}^{ij}_{12} = (1-p_{[2]})(\hat{\lambda}^{ij}_{23} + \hat{\lambda}^{ij}_{24}) +\hat{\lambda}^{ij}_{22}$;}
    \State{$\hat{\lambda}^{ij}_{11} = -\hat{\lambda}^{ij}_{12}$;}
   \State{\Return $\hat{\lambda}^{ij}_{k\ell}$ for $k=1,\dots,|\P_{ij}|$ and $\ell=1,2,3,4$.}
\end{algorithmic}
\end{algorithm}

\begin{myProp}[theorem style=plain]{}{prop_analytic}
Procedure $ComputeSol\lambda^{ij}$ provides an optimal solution of any dual slave problem $SP^{dual}_{ij}(v)$.
\end{myProp}
\begin{proof}
We consider the optimal solution \eqref{OptSPijS}-\eqref{OptSPijR} of problem $SP_{ij}(v)$ and apply the complementary slackness conditions from duality theory to deduce an optimal solution of problem $SP^{dual}_{ij}(v)$. At first, we consider the case where there is at least one variable $v_i = 1$ with $p_i = 0$ in the solution provided by the MP. This implies that nodes $i$ and $j$ are disconnected, i.e the optimal solution value of problem $SP_{ij}(v)$ is 0. Therefore a feasible and optimal dual solution is simply obtained by setting $\hat{\lambda}^{ij}_{k\ell} = 0$ for all $k$, $l$. 
Let us also recall that an optimal solution of MP sets $v_i = 0$ if $p_i = 1$. So without loss of generality we can consider MP solutions with $0 < p_i <1$ if $v_i = 1$ for $i=1, \dots, n$ for each path between two nodes $i$ and $j$. In these cases we have all the probabilities of survival $s^{ij}_k > 0$ for $k=1, \dots, |\P_{ij}|$, for each pair $(i,j)$.\\

\smallskip 

We proceed by considering the last node in the path, i.e. the value of $v_{[|\P_{ij}|]}$ and dual variables $\lambda^{ij}_{|\P_{ij}|\ell}$ $(\ell=1,\dots,4)$.
Since $s^{ij}_{|\P_{ij}|} > 0$, we have that an optimal dual solution must satisfy constraint \eqref{Dualslave4}  to equality, thus $\hat{\lambda}^{ij}_{|\P_{ij}|2} = c_{ij}$. To derive the values of $\lambda^{ij}_{|\P_{ij}|\ell}$ $(\ell = 1,3,4)$, we have to distinguish wheter $v_{|\P_{ij}|}$ is equal to 0 or 1. \\
If $v_{[|\P_{ij}|]} = 0$, constraints \eqref{slave4}-\eqref{slave5} (with $k=|\P_{ij}|$) are nonbinding. So the complementary slackness conditions indicate that dual variables $\lambda^{ij}_{|\P_{ij}|3}$ and $\lambda^{ij}_{|\P_{ij}|4}$ must be equal to 0, i.e. $\hat{\lambda}^{ij}_{|\P_{ij}|3} = \hat{\lambda}^{ij}_{|\P_{ij}|4} = 0$. Since variable $\lambda^{ij}_{|\P_{ij}|1}$ appears only in constraint \eqref{Dualslave2} and is not in the objective function anymore when $v_{[|\P_{ij}|]} = 0$, we can set $\hat{\lambda}^{ij}_{|\P_{ij}|1} = -\hat{\lambda}^{ij}_{|\P_{ij}|2}$ to satisfy constraint \eqref{Dualslave2}.\\
Consider now the case where $v_{[|\P_{ij}|]} = 1$. Primal constraint \eqref{slave3} is nonbinding implying $\hat{\lambda}^{ij}_{|\P_{ij}|1} = 0$ due to the associated complementary slackness condition. Also, from \eqref{OptSPijR} we have $r^{ij}_{|\P_{ij}|} > 0$ as  $p_{[|\P_{ij}|]} < 1$ and $s^{ij}_{|\P_{ij}|-1} > 0$. This implies that the corresponding dual constraint \eqref{Dualslave2} (with $k= |\P_{ij}|$) must be satisfied to equality in an optimal dual solution, i.e.:
$$\lambda^{ij}_{|\P_{ij}|1} + \lambda^{ij}_{|\P_{ij}|3} +\lambda^{ij}_{|\P_{ij}|4} + \lambda^{ij}_{|\P_{ij}|2} = 0 \implies \hat{\lambda}^{ij}_{|\P_{ij}|3} + \hat{\lambda}^{ij}_{|\P_{ij}|4} = -\hat{\lambda}^{ij}_{|\P_{ij}|2}.$$

Since the sum $\lambda^{ij}_{|\P_{ij}|3} + \lambda^{ij}_{|\P_{ij}|4}$ then appears in constraint \eqref{Dualslave3} with $k = |\P_{ij}|-1$ and variables $\lambda^{ij}_{|\P_{ij}|3}$, $\lambda^{ij}_{|\P_{ij}|4}$ are both excluded from the dual objective function when $v_{[|\P_{ij}|]} = 1$, we can arbitrarily set $\hat{\lambda}^{ij}_{|\P_{ij}|4} = 0$ and $\hat{\lambda}^{ij}_{|\P_{ij}|3} = -\hat{\lambda}^{ij}_{|\P_{ij}|2}$.
Notice that if we have $s^{ij}_{|\P_{ij}|-1} = 1$ when  $v_{[|\P_{ij}|]} = 1$, we can consider constraints \eqref{slave4} and \eqref{slave5} as nonbinding and equivalently set $\hat{\lambda}^{ij}_{|\P_{ij}|3} = \hat{\lambda}^{ij}_{|\P_{ij}|4} =0$, $\hat{\lambda}^{ij}_{|\P_{ij}|1} = -\hat{\lambda}^{ij}_{|\P_{ij}|2}$. Summarizing, the values of the dual variables associated with the last node in the path are:
$$ \hat{\lambda}^{ij}_{|\P_{ij}|1}=c_{ij}(v_{[|\P_{ij}|]}-1),\qquad \hat{\lambda}^{ij}_{|\P_{ij}|2}=c_{ij},\qquad \hat{\lambda}^{ij}_{|\P_{ij}|3}=-c_{ij}v_{[|\P_{ij}|]},\qquad \hat{\lambda}^{ij}_{|\P_{ij}|4}=0. $$

We now proceed by iteratively considering all the remaining nodes $k = |\P_{ij}|-1, \dots, 2$ and the value of the corresponding variables $v_{[k]}$. As $s^{ij}_{k} > 0$ for all $k$, the corresponding dual constraint \eqref{Dualslave3} must be satisfied to equality implying 

$$\hat{\lambda}^{ij}_{k2} = (1-p_{[k+1]})(\hat{\lambda}^{ij}_{(k+1)3} + \hat{\lambda}^{ij}_{(k+1)4}) +\hat{\lambda}^{ij}_{(k+1)2}.$$

After computing the value of $\hat{\lambda}^{ij}_{k2}$ with the last equation, we can apply the previous analysis to compute the values of $\hat{\lambda}^{ij}_{k1}$, $\hat{\lambda}^{ij}_{k3}$, $\hat{\lambda}^{ij}_{k4}$ according to the value taken by $v_{[k]}$. 
By using the previous reasonings for handling both the cases $v_{[k]}=0$ and $v_{[k]}=1$, we infer the following relations:

$$ \hat{\lambda}^{ij}_{k1}=(v_{[k]}-1)\hat{\lambda}^{ij}_{k2},\qquad \hat{\lambda}^{ij}_{k3}=-v_{[k]}\hat{\lambda}^{ij}_{k2},\qquad \hat{\lambda}^{ij}_{k4}=0. $$
When we reach the first node in the path, we have that dual constraint \eqref{Dualslave1} must be satisfied to equality as $s^{ij}_{1} > 0$. Hence we have $\hat{\lambda}^{ij}_{12} = (1-p_{[2]})(\hat{\lambda}^{ij}_{23} + \hat{\lambda}^{ij}_{24}) +\hat{\lambda}^{ij}_{22}$. Finally we can set $\hat{\lambda}^{ij}_{11} = -\hat{\lambda}^{ij}_{12}$ with both $v_{[1]} = 0$ and $v_{[1]} = 1$.
The computed solution is feasible as it satisfies all the dual constraints. We can show that such a solution is also optimal by strong duality.
If we recursively apply the relations identified by procedure $ComputeSol\lambda^{ij}$, we obtain the following expressions for each $\hat{\lambda}^{ij}_{k\ell}$ for $k=2, \dots, |\P_{ij}|$ and $\ell=1,2,3$:
$$ \hat{\lambda}^{ij}_{k1}=c_{ij}(v_{[k]}-1)\prod_{q>k}^{|\P_{ij}|}(1-(1-p_{[q]})v_{[q]}),\quad \hat{\lambda}^{ij}_{k2}=c_{ij}\prod_{q>k}^{|\P_{ij}|}(1-(1-p_{[q]})v_{[q]}),\quad \hat{\lambda}^{ij}_{k3}=-c_{ij}v_{[k]}\prod_{q>k}^{|\P_{ij}|}(1-(1-p_{[q]})v_{[q]}). $$
where the product $\prod_{q>k}^{|\P_{ij}|}(1-(1-p_{[q]})v_{[q]}$ is assumed to be 1 for $k = |\P_{ij}|$. For all $k > 1$ we have $\hat{\lambda}^{ij}_{k4} = 0$. For $k=1$ we have:
$$ \hat{\lambda}^{ij}_{11}=-c_{ij}\prod_{q>1}^{|\P_{ij}|}(1-(1-p_{[q]})v_{[q]}),\qquad \hat{\lambda}^{ij}_{12}=c_{ij}\prod_{q>1}^{|\P_{ij}|}(1-(1-p_{[q]})v_{[q]}). $$
When plugging these values inside the dual objective function \eqref{DualFOslave}, we notice that the last sum $\sum_{k=2}^{|\P_{ij}|} ((1-p_{[k]})v_{[k]}\lambda^{ij}_{k1} + (1-p_{[k]})(v_{[k]}-1)\lambda^{ij}_{k4})$ is 0 as $\lambda^{ij}_{k4}$ is 0 and there are factors $v_{[k]}(1-v_{[k]})=0$ induced by the value of $\lambda^{ij}_{k1}$. The first two terms instead sum up to $c_{ij}\prod_{q\geq1}^{|\P_{ij}|}(1-(1-p_{[q]})v_{[q]})$. Since this value coincides with the optimal value of the primal problem $SP_{ij}(v)$, the dual solution considered is optimal.
\end{proof}

\section{Computational results}
\label{sec:numerical}

We first describe the setting of our numerical experiments as well as the characteristics of the different instances generated. Then we discuss the results and the performance of the proposed approaches.

\subsection{Experimental conditions and instances}

All algorithms have been implemented in C++, compiled with \texttt{gcc~4.1.2}, and all tests were performed on an HP ProLiant DL585 G6 server with two 2.1 GHz AMD Opteron 8425HE processors (with 12 threads) and 24 GB of RAM. We used solver CPLEX 12.8 to solve the MILP formulation \eqref{Eq:THEObj}-\eqref{Eq:THEVarV} and the MP along the iterations of algorithm \BDAL. For the MILP formulation, we set the precision level (absolute gap) of  CPLEX 12.8 to 0.001. The other parameters of the solver were set to their default values. In algorithm \BDAL we set $\varepsilon = 0.001$ for a fair comparison with CPLEX 12.8 launched on the MILP formulation. We generated trees with the number of nodes $n$ ranging from 20 to 100. For each value of $n$, 30 different trees were generated as in~\cite{DisGroLoc11cnp}, i.e. using Broder’s algorithm for the uniform spanning tree problem~\cite{Broder1989}, which guarantees that each tree is randomly chosen with uniform distribution among all trees with $n$ nodes. For each instance we generated a set of probabilities of survival $p_i$ ($i\in V$) uniformly distributed between 0 and 1 and at most two decimal digits. 
For each tree we considered different sets of weights. A first set of \textit{unweighted instances} considers both unit attack costs $\kappa_i$ and unit connection costs $c_{ij}$. Then, we generated \textit{weighted instances} with three different sets of weights. A first set of weights (\textit{type 1}) considers both attack costs and connection costs as integers uniformly distributed in the interval $[1, 10]$. A second set of weights (\textit{type 2}) still considers connections costs $c_{ij} \in[1,10]$
but attack costs $\kappa_i \in[1,100]$. The last set of weights  (\textit{type 3}) considers connection costs $c_{ij} \in[1,10]$
and attack costs equal to the inverse of the survival probability of each node, i.e. $\kappa_i=1/p_i$ for each $i\in V$, where we set $\kappa_i=100$ if $p_i=0$. For this last set of weights, the rationale is to generate instances which are expected to be hard to solve in the sense that nodes with a lower attack cost have also a higher probability of survival. Summarizing, each instance has a unique set of probabilities $p_i$ but four different sets of attack costs and connection costs. The budget $K$ is equal to 10\% of the sum of the attack costs of the nodes: $K=0.1\sum_{i\in V}\kappa_i$. We remark that even for the smallest instances with $n=20$, the use of the general model \eqref{Eq:Obj1}-\eqref{Eq:varVi1} would be impractical due to the induced huge number of variables and constraints given by the number of scenarios to analyze. Our approach does not explicitly require the definition of all possible scenarios and thus can be applied to larger instances within reasonable computational times, as discussed in the next sections. We tested the proposed MILP formulation of the problem and algorithm \BDAL with a time limit of 3600 seconds. 

\subsection{Results for the unweighted instances}

We first present the computational results for the unweighted instances in Table~\ref{tab:unit}. For each given number of nodes $n$, the table reports the performance CPLEX 12.8 launched on the MILP formulation over 30 instances in terms of average computational time (column \textit{time}) and average percentage gap $1 - \frac{LB}{UB}$ (column \textit{gap (\%)})  between the best solution found ($UB$) for the problem and its lower bound ($LB$). The number of the instances solved within the time limit is also reported (column \textit{\# closed}). The average values consider also the instances where the time limit is reached. The same quantities are reported for algorithm \BDAL with two additional columns: the average number of iterations (column \textit{\# iter.}) executed by the algorithm and the average number of cuts added over all iterations (column \textit{\# cuts}). The best percentage gap between the two approaches is displayed in bold font. 
The results in Table~\ref{tab:unit} illustrate that CPLEX 12.8 launched on the MILP formulation \eqref{Eq:THEObj}-\eqref{Eq:THEVarV} successfully solves instances with up to 60 nodes and all the instances with 80 nodes but one. For larger instances with up to 150 nodes, the solver hardly succeeds in solving the instances within the time limit of 3600 seconds and in obtaining relatively small percentage gaps. 

Algorithm \BDAL shows up to outperform CPLEX 12.8. The proposed algorithm solves all 30 instances for trees with up to 80 nodes with a lower average running time except for the smallest instances with 20 nodes. For larger instances, the running times and percentage gaps provided by \BDAL are considerably smaller than the ones given by CPLEX 12.8.   
Even on instances with 150 nodes, algorithm \BDAL manages to close more than two third of the benchmark instances while CPLEX 12.8 closes none of the 30 instances. We point out that in our numerical experiments the valid inequalities of Proposition \ref{th:prop:valid_ineq} tend to speed up the performance of algorithm \BDAL by a sizeable factor on instances with up to 100 nodes, providing a reduction by 20-30\% in the running times. This effect is harder to evaluate in the instances where the algorithm runs out of time.

\begin{table}[H]
\centering
\begin{tabular}{lrrrrrrrrr}
\sphline{3}
$n$ & \multicolumn{3}{c}{CPLEX 12.8} && \multicolumn{5}{c}{\BDAL} \\
 & time & gap (\%) & \# closed && time & gap (\%) & \# closed & \# iter. & \# cuts \\
\noalign{\vskip 2 pt}\cline{2-4}\cline{6-10}\noalign{\vskip 3 pt}
20  &  0.40    & \textbf{0.00}  & 30 &&  0.90 & \textbf{0.00} & 30 & 2.47 & 279.60 \\
40  &  9.10    & \textbf{0.00}  & 30 &&  2.43 & \textbf{0.00} & 30 & 3.73 & 1717.73 \\
60  &  215.10  & \textbf{0.00}  & 30 &&  11.17 & \textbf{0.00} & 30 & 4.87 & 4930.23 \\
80  &  1276.87 & 0.16           & 29 &&  57.83 & \textbf{0.00} & 30 & 5.47 & 10446.43 \\
100 &  3232.73 & 5.29           & 11 &&  603.87 & \textbf{0.06} & 28 & 7.67 & 21822.43 \\
120 &  3588.00 & 48.23          & 1  &&  976.97 & \textbf{0.06} & 28 & 7.27 & 29918.90 \\
150 &  3600.00 & 86.32          & 0  &&  1894.37 & \textbf{0.47} & 21 & 7.00 & 47984.67 \\
\sphline{3}
\end{tabular}
\caption{\Trees unweighted instances.}
\label{tab:unit}
\end{table}

\subsection{Results for the weighted instances}

The results for the three types of unweighted instances are reported in Tables~\ref{tab:weights1}-\ref{tab:weights3}. For CPLEX 12.8 we report the results without the valid inequalities of Proposition \ref{th:prop:valid_ineq} in the MILP model, since the solver performs better without these additional constraints in the weighted instances. Instead, the valid inequalities are still beneficial for algorithm \BDAL as for the unweighted instances. The comparison between CPLEX 12.8 and algorithm \BDAL confirms the previous trend in the results. In general the proposed algorithm solves more instances than CPLEX 12.8 within the considered time limit. For large instances with more than 80 nodes, CPLEX 12.8 is not capable of solving the instances with a reasonably small percentage gap, while algorithm \BDAL provides average gaps lower than 1\%, except for the largest instances of type 3. As we should expect, the type 3 instances turn out to be the hardest instances to solve. Nevertheless, algorithm \BDAL provides solutions for the largest type 3 instances with 150 nodes with an average percentage gap of 6\%, although the algorithm does not close any instance within the time limit. With respect to the unweighted instances, we notice that the number of iterations and the number of generated cuts of the algorithm tend to increase for weighted instances, in particular for type 3 instances.

\begin{table}[H]
\centering
\begin{tabular}{lrrrrrrrrr}
\sphline{3}
$n$ & \multicolumn{3}{c}{CPLEX 12.8} && \multicolumn{5}{c}{\BDAL} \\
 & time & gap (\%) & \# closed && time & gap (\%) & \# closed & \# iter. & \# cuts \\
\noalign{\vskip 2 pt}\cline{2-4}\cline{6-10}\noalign{\vskip 3 pt}
20  &  2.77    & \textbf{0.00}  & 30 &&  2.47    & \textbf{0.00} & 30 & 2.57 & 314.70 \\
40  &  22.93   & \textbf{0.00}  & 30 &&  7.67    & \textbf{0.00} & 30 & 5.33 & 2408.50 \\
60  &  192.80  & \textbf{0.00}  & 30 &&  40.03   & \textbf{0.00} & 30 & 6.23 & 6039.70 \\
80  &  1454.50 & 0.10           & 29 &&  137.70  & \textbf{0.00} & 30 & 7.00 & 12817.87 \\
100 &  3245.00 & 4.28           & 10 &&  600.00  & \textbf{0.12} & 29 & 7.97 & 23527.83 \\
120 &  3600.00 & 37.61          & 0  &&  1393.87 & \textbf{0.04} & 24 & 8.87 & 35284.70 \\
150 &  3600.00 & 78.60          & 0  &&  2607.03 & \textbf{0.78} & 16 & 7.53 & 53327.30 \\
\sphline{3}
\end{tabular}
\caption{\Trees weighted instances (type 1).} 
\label{tab:weights1}
\end{table}

\begin{table}[H]
\centering
\begin{tabular}{lrrrrrrrrr}
\sphline{3}
$n$ & \multicolumn{3}{c}{CPLEX 12.8} && \multicolumn{5}{c}{\BDAL} \\
 & time & gap (\%) & \# closed && time & gap (\%) & \# closed & \# iter. & \# cuts \\
\noalign{\vskip 2 pt}\cline{2-4}\cline{6-10}\noalign{\vskip 3 pt}
20  &  2.07     & \textbf{0.00} & 30 &&  4.00    & \textbf{0.00} & 30 & 2.87 & 366.23 \\
40  &  16.43    & \textbf{0.00} & 30 &&  5.43    & \textbf{0.00} & 30 & 4.53 & 2134.83 \\
60  &  241.23   & \textbf{0.00} & 30 &&  29.47   & \textbf{0.00} & 30 & 5.13 & 5555.47 \\
80  &  1163.43  & \textbf{0.00} & 30 &&  92.20   & \textbf{0.00} & 30 & 6.47 & 11915.43 \\
100 &  3229.83  & 4.05          & 12 &&  596.20  & \textbf{0.01} & 29 & 8.37 & 22930.93 \\
120 &  3600.00  & 37.88         & 0  &&  1125.20 & \textbf{0.06} & 26 & 7.60 & 31621.47 \\
150 &  3600.00  & 81.01         & 0  &&  2671.90 & \textbf{0.60} & 16 & 7.80 & 53516.63 \\
\sphline{3}
\end{tabular}
\caption{\Trees weighted instances (type 2).} 
\label{tab:weights2}
\end{table}

\begin{table}[H]
\centering
\begin{tabular}{lrrrrrrrrr}
\sphline{3}
$n$ & \multicolumn{3}{c}{CPLEX 12.8} && \multicolumn{5}{c}{\BDAL} \\
 & time & gap (\%) & \# closed && time & gap (\%) & \# closed & \# iter. & \# cuts \\
\noalign{\vskip 2 pt}\cline{2-4}\cline{6-10}\noalign{\vskip 3 pt}
20  &  2.93     & \textbf{0.00} & 30 &&  6.90    & \textbf{0.00} & 30 & 5.27  & 608.37 \\
40  &  19.67    & \textbf{0.00} & 30 &&  32.57   & \textbf{0.00} & 30 & 8.17  & 3723.57 \\
60  &  376.33   & \textbf{0.00} & 30 &&  321.37  & \textbf{0.00} & 30 & 10.83 & 11301.63 \\
80  &  1852.03  & 0.52          & 26 &&  1181.67 & \textbf{0.06} & 27 & 10.67 & 19941.83 \\
100 &  3453.67  & 7.43          & 4  &&  2327.87 & \textbf{0.56} & 18 & 9.80  & 32258.57 \\
120 &  3600.00  & 26.19         & 0  &&  3270.50 & \textbf{2.91} & 5  & 7.87  & 42625.43 \\
150 &  3600.00  & 66.78         & 0  &&  3600.00 & \textbf{6.15} & 0  & 5.70  & 56972.43 \\
\sphline{3}
\end{tabular}
\caption{\Trees weighted instances (type 3).} 
\label{tab:weights3}
\end{table}

\section{Conclusions}
\label{sec:conclusion}

In this work we introduced a stochastic version of the Critical Node Problem where the outcome of attacks on nodes is uncertain. 
We first presented an Integer Linear Programming formulation for the problem over general graphs and derived valid inequalities. Then we focused on the problem variant over trees for which we provided theoretical results, a nonlinear model and a linear reformulation based on the probability chains from probability theory. 
Moreover, we proposed a dedicated algorithm to solve the derived linearized model through a Benders Decomposition approach. 
We performed an extensive computational analysis to assess the effectiveness and robustness of the proposed approaches. 
We also introduced an ILP formulation and an approximation algorithm based on a dynamic program for specific problem variants of interest.

Numerous research lines could be explored in the future. 
The most natural follow-up of our work would be the design of a solution approach for tackling the stochastic CNP over general graphs. A possibility is to combine modern scenario partitioning methods such as in~\cite{Cormican1998,Song2015,vanAckooij2018} with a Column Generation approach.
It might be also interesting to adapt the proposed algorithmic framework based on Benders Decomposition to the deterministic version of the CNP on general graphs. Finally, exploring the design of efficient heuristic/metaheuristic algorithms could be a valid alternative to solve instances on large (and arbitrary) graphs. 
This last option would call for finding efficient procedures to compute the stochastic objective function of a given solution, since the complexity of computing this value is \emph{a priori} exponential in the 
solution size.
For example, we could derive a FPRAS to compute the objective function with a suitable precision, as it was done, e.g., in~\cite{stoch-CNP_Thai2015} for the stochastic CNP with edge uncertainty.

\section*{Acknowledgements}
We would like to thank the authors of~\cite{DisGroLoc11cnp} 
for sharing their code for the generation of the instances. We are in debt with Konstantin Pavlikov for bringing the Chain Rule method to our attention.

\bibliographystyle{plain}
\bibliography{CNP}

\begin{thebibliography}{10}

\bibitem{Addis13}
B.~Addis, M.~{Di~Summa}, and A.~Grosso.
\newblock Removing critical nodes from a graph: complexity results and
  polynomial algorithms for the case of bounded treewidth.
\newblock {\em Discrete Applied Mathematics}, 16-17:2349--2360, 2013.

\bibitem{INOC2015}
R.~Aringhieri, A.~Grosso, and P.~Hosteins.
\newblock A {G}enetic {A}lgorithm for a class of {C}ritical {N}ode {P}roblems.
\newblock In {\em The 7th International Network Optimization Conference
  (INOC'15)}, volume~52 of {\em Electronic Notes in Discrete Mathematics},
  pages 359--366, 2016.

\bibitem{Genetic}
R.~Aringhieri, A.~Grosso, P.~Hosteins, and R.~Scatamacchia.
\newblock A general {E}volutionary {F}ramework for different classes of
  {C}ritical {N}ode {P}roblems.
\newblock {\em Engineering Applications of Artificial Intelligence},
  55:128--145, 2016.

\bibitem{Networks}
R.~Aringhieri, A.~Grosso, P.~Hosteins, and R.~Scatamacchia.
\newblock Local {S}earch {M}etaheuristics for the {C}ritical {N}ode {P}roblem.
\newblock {\em Networks}, 67(3):209--221, 2016.

\bibitem{DAM-DCNP}
R.~Aringhieri, A.~Grosso, P.~Hosteins, and R.~Scatamacchia.
\newblock Polynomial and pseudo-polynomial time algorithms for different
  classes of the {D}istance {C}ritical {N}ode {P}roblem.
\newblock {\em Discrete Applied Mathematics}, 253:103--121, 2019.

\bibitem{Aru-al09cnp}
A.~Arulselvan, C.~W. Commander, L.~Elefteriadou, and P.~M. Pardalos.
\newblock Detecting critical nodes in sparse graphs.
\newblock {\em Computers \& Operations Research}, 36:2193--2200, 2009.

\bibitem{Ball1989}
M.~O. Ball, B.~L. Golden, and R.~V. Vohra.
\newblock Finding the most vital arcs in a network.
\newblock {\em Operations Research Letters}, 8:73--76, 1989.

\bibitem{Benders62}
J.F. Benders.
\newblock Partitioning procedures for solving mixed-variables programming
  problems.
\newblock {\em Numerische Mathematik}, 4:238--252, 1962.

\bibitem{Broder1989}
A.~Broder.
\newblock Generating random spanning trees.
\newblock In {\em In: 30th annual symposium on foundations of computer
  science}, pages 442--447, 1989.

\bibitem{homelandsec}
Gerald Brown, Matthew Carlyle, Javier Salmer\'on, and Kevin Wood.
\newblock Defending critical infrastructure.
\newblock {\em Interfaces}, 36(6):530--544, 2006.

\bibitem{Corley1974}
H.~W. Corley and H.~Chang.
\newblock Finding the n most vital nodes in a flow network.
\newblock {\em Management Science}, 21(3):362--364, 1974.

\bibitem{Corley1982}
H.W. Corley and D.~Y. Sha.
\newblock Most vital links and nodes in weighted networks.
\newblock {\em Operations Research Letters}, 1(4):157 -- 160, 1982.

\bibitem{Cormican1998}
K.~J. Cormican, D.~P. Morton, and R.~K. Wood.
\newblock Stochastic network interdiction.
\newblock {\em Operations Research}, 46:184--197, 1998.

\bibitem{DisGroLoc11cnp}
M.~Di~Summa, A.~Grosso, and M.~Locatelli.
\newblock The critical node problem over trees.
\newblock {\em Computers and Operations Research}, 38:1766--1774, 2011.

\bibitem{DiSumma2012}
M.~{Di~Summa}, A.~Grosso, and M.~Locatelli.
\newblock Branch and cut algorithms for detecting critical nodes in undirected
  graphs.
\newblock {\em Computational Optimization and Applications}, 53:649--680, 2012.

\bibitem{stoch-CNP_Thai2015}
T.~N. Dinh and M.~T. Thai.
\newblock Assessing attack vulnerability in networks with uncertainty.
\newblock In {\em 2015 IEEE Conference on Computer Communications (INFOCOM)},
  pages 2380--2388, 2015.

\bibitem{PardalosRobust}
N.~Fan and P.~M. Pardalos.
\newblock Robust optimization of graph partitioning and critical node detection
  in analyzing networks.
\newblock In W.~Wu and O.~Daescu, editors, {\em Combinatorial Optimization and
  Applications}, pages 170--183, Berlin, Heidelberg, 2010. Springer Berlin
  Heidelberg.

\bibitem{Granata2013}
D.~Granata, G.~Steeger, and S.~Rebennack.
\newblock Network interdiction via a critical disruption path: Branch-and-price
  algorithms.
\newblock {\em Computers \& Operations Research}, 40(11):2689 -- 2702, 2013.

\bibitem{Hooshmand2019}
F.~Hooshmand, F.~Mirarabrazi, and S.A. MirHassani.
\newblock Efficient benders decomposition for distance-based critical node
  detection problem.
\newblock {\em Omega}, 2019.

\bibitem{Israeli2002}
E.~Israeli and R.~K. Wood.
\newblock Shortest-path network interdiction.
\newblock {\em Networks}, 40(2):97--111, 2002.

\bibitem{Jenelius2006537}
Erik Jenelius, Tom Petersen, and Lars-Göran Mattsson.
\newblock Importance and exposure in road network vulnerability analysis.
\newblock {\em Transportation Research Part A: Policy and Practice}, 40(7):537
  -- 560, 2006.

\bibitem{Lalou2016}
M.~Lalou, M.~A. Tahraoui, and H.~Kheddouci.
\newblock Component-cardinality-constrained critical node problem in graphs.
\newblock {\em Discrete Applied Mathematics}, 210:150--163, 2016.

\bibitem{Lalou2018}
M.~Lalou, M.~A. Tahraoui, and H.~Kheddouci.
\newblock The critical node detection problem in networks: A survey.
\newblock {\em Computer Science Review}, 28:92--117, 2018.

\bibitem{IJoCRobust}
J.~Naoum-Sawaya and C.~Buchheim.
\newblock Robust critical node selection by benders decomposition.
\newblock {\em INFORMS Journal on Computing}, 28(1):162--174, 2016.

\bibitem{OHanley2013}
Jesse~R. O’Hanley, M.~Paola Scaparra, and Sergio García.
\newblock Probability chains: A general linearization technique for modeling
  reliability in facility location and related problems.
\newblock {\em European Journal of Operational Research}, 230(1):63 -- 75,
  2013.

\bibitem{Pavlikov2018}
Konstantin Pavlikov.
\newblock Improved formulations for minimum connectivity network interdiction
  problems.
\newblock {\em Computers \& Operations Research}, 97:48 -- 57, 2018.

\bibitem{GRASP-CNP2}
D.~Purevsuren, G.~Cui, M.~Chu, and N.~N.~Htay Win.
\newblock Hybridization of grasp with exterior path relinking for identifying
  critical nodes in graphs.
\newblock {\em IAENG International Journal of Computer Science}, 44, 2017.

\bibitem{RaCrGeRe17}
R.~Rahmanian, T.G. Crainic, M.~Gendreau, and W.~Rei.
\newblock The {B}enders decomposition algorithm: A literature review.
\newblock {\em European Journal of Operational Research}, 259:801--817, 2017.

\bibitem{Salmeron2004}
J.~Salmer\'on, K.~Wood, and R.~Baldick.
\newblock Analysis of electric grid security under terrorist threat.
\newblock {\em IEEE Trans Power Syst}, 19:905--912, 2004.

\bibitem{Seo2015}
J.~Seo, S.~Mishra, X.~Li, and M.~T. Thai.
\newblock Catastrophic cascading failures in power networks.
\newblock {\em Theoretical Computer Science}, 607:306--319, 2015.

\bibitem{SmithShen12}
S.~Shen and J.~C. Smith.
\newblock Polynomial-time algorithms for solving a class of critical node
  problems on trees and series-parallel graphs.
\newblock {\em Networks}, 60(2):103--119, 2012.

\bibitem{Smith2013}
J.~C. Smith, M.~Prince, and J.~Geunes.
\newblock {\em Modern Network Interdiction Problems and Algorithms}, pages
  1949--1987.
\newblock Springer New York, New York, NY, 2013.

\bibitem{Song2015}
Y.~Song and J.~Luedtke.
\newblock An adaptive partition-based approach for solving two-stage stochastic
  programs with fixed recourse.
\newblock {\em SIAM Journal on Optimization}, 25(3):1344--1367, 2015.

\bibitem{Tomaino2012}
Vera Tomaino, Ashwin Arulselvan, Pierangelo Veltri, and Panos~M. Pardalos.
\newblock {\em Studying Connectivity Properties in Human Protein--Protein
  Interaction Network in Cancer Pathway}, pages 187--197.
\newblock Springer US, Boston, MA, 2012.

\bibitem{vanAckooij2018}
W.~van Ackooij, W.~de~Oliveira, and Y.~Song.
\newblock Adaptive partition-based level decomposition methods for solving
  two-stage stochastic programs with fixed recourse.
\newblock {\em INFORMS Journal on Computing}, 30(1):57--70, 2018.

\bibitem{Ventresca2014b}
M.~Ventresca and D.~Aleman.
\newblock A randomized algorithm with local search for containment of pandemic
  disease spread.
\newblock {\em Computers \& Operations Research}, 48:11--19, 2014.

\bibitem{Veremyev2014}
A.~Veremyev, V.~Boginski, and E.~Pasiliao.
\newblock Exact identification of critical nodes in sparse networks via new
  compact formulations.
\newblock {\em Optimization Letters}, 8:1245--1259, 2014.

\bibitem{Veremyev2014b}
A.~Veremyev, O.~A. Prokopyev, and E.~L. Pasiliao.
\newblock An integer programming framework for critical elements detection in
  graphs.
\newblock {\em Journal of Combinatorial Optimization}, 28:233--273, 2014.

\bibitem{Veremyev2015}
A.~Veremyev, O.~A. Prokopyev, and E.~L. Pasiliao.
\newblock Critical nodes for distance-based connectivity and related problems
  in graphs.
\newblock {\em Networks}, 66:170--195, 2015.

\bibitem{Walteros2018}
Jose~L. Walteros, Alexander Veremyev, Panos~M. Pardalos, and Eduardo~L.
  Pasiliao.
\newblock Detecting critical node structures on graphs: A mathematical
  programming approach.
\newblock {\em Networks}, 2018.

\bibitem{Wood93}
R.~K. Wood.
\newblock Deterministic network interdiction.
\newblock {\em Mathematical and Computer Modelling}, 17:1--18, 1993.

\bibitem{Zhou2006}
T.~Zhou, Z.~Q. Fu, and B.~H. Wang.
\newblock Epidemic dynamics on complex networks.
\newblock {\em Progress in Natural Sciences}, 16:452--457, 2006.

\bibitem{Glover2018}
Y.~Zhou, J.~Hao, and F.~Glover.
\newblock Memetic search for identifying critical nodes in sparse graphs.
\newblock {\em IEEE Transactions on Cybernetics}, 2018.

\end{thebibliography}

\begin{appendices}
\section{A linear model with equal survival probabilities}
\label{app:ILP_equal_p}

For the simplified problem variant where all survival probabilities are equal, i.e $p_i=p$ for $i\in V$, we derive an ILP formulation that implicitly takes into account the uncertainty associated with the node attacks. 

We observe that for this problem we only need to know the number of nodes attacked in $\P_{ij}$ to compute the average connection cost $c_{ij}\prod_{k\in S_{ij}}p_k=c_{ij}p^{|S_{ij}|}$ 
between nodes $i$ and $j$. We can thus introduce binary variables $y_r^{(ij)}$, equal to 1 only if there are $r$ nodes attacked in path $\P_{ij}$. Index $r$ can vary from 0 to $\rho_{ij}$, with $\rho_{ij} := \min \{\frac{K}{\min_{i\in V}  \kappa_i}, |\P_{ij}| \}$. We denote by ILP$_p$ the following ILP formulation:
\begin{subequations}
\label{Eq:same_p}
\begin{align}
\min & \sum_{i<j}c_{ij}\sum_{r \leq \rho_{ij}}p^ry_{r}^{(ij)}		 \label{Eq:PolObj}\\
     & \sum_i \kappa_iv_i \leq K                                                   	\label{Eq:BudSec} \\
     & \sum_{r \leq \rho_{ij}}y_{r}^{(ij)} = 1							& i,j\in V: i<j \label{Eq:PolConst1}\\
     & \sum_{r \leq \rho_{ij}}ry_{r}^{(ij)} = \sum_{k\in\P_{ij}}v_k				& i,j\in V: i<j \label{Eq:PolConst2}\\
     & y_{r}^{(ij)} \in\{0,1\}  & i,j\in V: i<j, r = 0, \dots, \rho_{ij} \label{Eq:varUijR} \\
     & v_i\in\{0,1\} & i\in  V  \label{Eq:varViSec}
\end{align}
\end{subequations}
The objective function \eqref{Eq:PolObj} minimizes the expected cost of the pairwise connectivity after deletion of nodes in the tree. Constraint \eqref{Eq:BudSec} represents the budget constraint. Constraints \eqref{Eq:PolConst1} ensure that only one value of $r$ will be selected for each path $\P_{ij}$. Constraints \eqref{Eq:PolConst2} links $y_{r}^{(ij)}$ variables with $v_i$ variables. Constraints \eqref{Eq:varUijR} and \eqref{Eq:varViSec} define the domain of the variables. Even though the number of variables $y_{r}^{(ij)}$ is increased by a factor $\rho_{ij}$ with respect to the corresponding $u_{ij}$  variables in the deterministic CNP, model  \eqref{Eq:PolObj}-\eqref{Eq:varViSec} can constitute a valid alternative of practical relevance. Preliminary computational tests illustrate that solving this problem variant by model ILP$_p$ is more effective than using the general model \eqref{Eq:THEObj}-\eqref{Eq:THEVarV} and the proposed exact approach.

\section{An approximation algorithm for the Stochastic CNP over trees}
\label{app:approx}

As an additional result, we present an approximation algorithm with an absolute performance guarantee for the \Trees 
with unit connection costs and unit attack costs. 
The algorithm relies on limiting the numerical precision for the evaluation of each term in the objective function. We remark that the algorithm is very similar to the one we proposed in~\cite{DAM-DCNP} for a certain class of multiplicative distance function for the Distance CNP over trees. Our goal here is to show how to apply similar techniques to derive an approximation algorithm for the considered stochastic problem. For the sake of conciseness, we will refer the reader to our previous work for the details of the algorithm and recursions. 
We first sketch a Dynamic Programming (DP) algorithm for the problem which constitutes the basis for the approximation algorithm. To derive a DP algorithm for the \Trees, 
we will require that each connection cost in the objective can be scaled to an integer in the recursions. 
To this aim, we will consider the minimum power of 10, denoted by $\mu$, such that $\mu \prod_{k\in\P_{ij}}p_k\in\mathbb{N}$ for any node pair $(i,j)\in V\times V$. Note that the value of parameter $\mu$ could be exponentially large depending on the value of probabilities $p_k$, which could compromise the performance of an algorithm based on scaling the objective to an integer value. However, as shown later on the proposed approximation algorithm is based on limiting $\mu$ to reasonable values.

Let us denote by $\T_a$ the subtree of tree $\T$ rooted at node $a\in V$, 
and by $a_i$ with $i\in\{1,...,s\}$ the children of $a$. 
Also, we define as $\T_{a_{i\rightarrow s}}$ the subtree constituted
by $\{a\}\cup_{j=i,...,s}\T_{a_j}$. 
For further details about this particular recursion scheme we refer the interested reader to~\cite{DAM-DCNP,DisGroLoc11cnp}.
All recursions in the dynamic programming approach are based on traversing the tree in postorder
(i.e.\ from the leaves to the root) and from the right part of each tree level to the left one.
We define the following recursion functions:

\begin{parcolumns}[colwidths={1=.2\textwidth},distance=1em]{2}
\colchunk{
\hfill $F_a(c,k,\sigma)$ \hfill :=
}
\colchunk{%
minimum cost of a solution for subtree $\T_a$ when $k$ nodes are deleted 
from $\T_a$ and the total cost of connecting $a$ to subtree $\T_a$ multiplied by $\mu$ is $c$. Index $\sigma$ is equal to either 1 if $a\in S$ or 0 if $a\notin S$. 
}
\end{parcolumns}

\begin{parcolumns}[colwidths={1=.2\textwidth},distance=1em]{2}
\colchunk{%
\hfill $G_{a_i}(c,k,\sigma)$ \hfill :=
}
\colchunk{%
minimum cost of a solution for subtree $\T_{a_{i\rightarrow s}}$ when $k$ nodes are 
deleted from $\T_{a_{i\rightarrow s}}$ and the total cost of connecting $a$ to subtree 
$\T_{a_{i\rightarrow s}}$ multiplied by $\mu$ is $c$. Index $\sigma$ is equal to either 1 if $a\in S$ or 0 if $a\notin S$.
}
\end{parcolumns}
If it is not possible to remove $k$ nodes from $\T_a$ such that the total cost of connecting $a$ to subtree $\T_a$ multiplied by $\mu$ is equal to $c$, we have $F_a(c,k,\sigma)=\infty$ and/or $G_a(c,k,\sigma) =\infty$.

Denoting by $C(\T')$ the connectivity cost of any subtree $\T'$, we derive the following recursions for non-leaf node $a\in V$:
\begin{dmath}
F_a(c,k,\sigma) = G_{a_1}(c,k,\sigma);
\label{eq:mult_rec1}
\end{dmath}
for $i<s$, if $p_a=0$ and $a\in S$:
\begin{dgroup}
\begin{dmath}
G_{a_i}(c,k,1) = \min \left\{
        F_{a_i}(b,q,\sigma^\prime)+G_{a_{i+1}}\left(0,k-q,1\right) \\
	b \hiderel{=}0,\dots,\mu|\T_{a_i}|;\, 
	q \hiderel{=}0,\dots,k-1;\,
	\sigma^\prime \hiderel{=}0,1
\right\},
\end{dmath}
\begin{dsuspend}
otherwise, if $a \notin S$ or $p_a>0$:
\end{dsuspend}
\begin{dmath}
G_{a_i}(c,k,\sigma) = \min \left\{
	F_{a_i}(b,q,\sigma^\prime)+G_{a_{i+1}}\left(c-p_a^{\sigma}(b+\mu p_{a_{i+1}}^{\sigma^\prime}),k-q,\sigma\right) \\
	\text{$+$ } p_a^\sigma p_{a_i}^{\sigma^\prime} + p_{a_i}^{\sigma^\prime}\left(\dfrac{c}{\mu}-p_a^{\sigma}\left(\dfrac{b}{\mu}+p_{a_{i+1}}^{\sigma^\prime}\right)\right) + p_a^\sigma \frac{b}{\mu} + \dfrac{b}{\mu}\left(\dfrac{c}{\mu}-p_a^{\sigma}\left(\dfrac{b}{\mu}+p_{a_{i+1}}^{\sigma^\prime}\right)\right)\\
	b \hiderel{=}0,\dots,p_a^{-\sigma}c-\mu p_{a_i}^{\sigma^\prime};\, 
	q \hiderel{=}0,\dots,k-\sigma;\,
	\sigma^\prime \hiderel{=}0,1
\right\}.
\label{eq:mult_rec2}
\end{dmath}
\end{dgroup}
The second line in 
Equation~\eqref{eq:mult_rec2} represents the total cost of connecting subtrees $\T_{a_i}$ 
and $\T_{a_{i\rightarrow s}}$. The first term in this line represents the connection cost of node $a$ to $a_i$, the second term represents the connection of $a_i$ with all the nodes in $\T_{a_{i\rightarrow s}}$ and the third term the connection of $a$ with all the nodes in $\T_{a_i}$. Finally, the last term represents the connection cost between each node 
$u\in \T_{a_i}\setminus\{a_i\}$ and $v\in \T_{a_{i\rightarrow s}}\setminus\{a\}$. The cost for connecting pair $(u,v)$ 
is in fact given by $E_{s\in\S}[u^s_{uv}]=E_{s\in\S}[u^s_{ua_i}]E_{s\in\S}[u^s_{av}]$ where $E_{s\in\S}[u^s_{uv}]$ is the average value of the connection between nodes $u$ and $v$. Summing over all possible node pairs $(u,v)$ gives the last term in Equation~\eqref{eq:mult_rec2}. Note that if $p_a=0$ and $\sigma=0$, we assume in our equations that $p_a^{\pm\sigma}=1$.

We do not provide extensive equations for the initial conditions as they are in fact the same as those provided in~\cite{DAM-DCNP}. 
The optimal solution value is given by  
$$\min\left\{F_r(c,k,\sigma):\, \,c=0,\dots,C(\T);\,k=0,\dots,K;\,\sigma=0,1 \right\}$$ 
where $r$ is the root node of the tree. The optimal solution set of deleted nodes can be recovered by backtracking. Using the previous DP algorithm, we can state the following proposition.

\begin{myProp}[theorem style=plain]{}{propertyC3-1}
The stochastic CNP over trees with unit pair costs and unit deletion costs can be solved with a time complexity of $\O(K^2n^2\mu^2)$.
\end{myProp}
\begin{proof}
The number of all possible combinations of the indices in functions $F_a$ and $G_a$ is bounded by $2[(n-1)\mu+1](K+1)$. The recursion step with the largest number of operations is given by Equation~\eqref{eq:mult_rec2} which requires up to $2[(n-1)\mu+1](K+1)$ operations. Thus the running time of the DP algorithm can be bounded by $\O(K^2n^2\mu^2)$.
\end{proof}

Given the complexity stated in the above proposition, the performance of the proposed dynamic program is heavily affected by large values of $\mu$. However, the DP algorithm can be modified to devise a more practical approximation algorithm, which also provides a lower bound on the optimal objective value. Because of the rapid decrease of the cost function with the number of multiplying probabilities, beyond a certain number of deleted nodes the connection cost between two nodes will only increase the objective function by a negligible amount. Therefore, we propose to modify the algorithm presented in Equations~\eqref{eq:mult_rec1}-\eqref{eq:mult_rec2} by truncating each term in the objective after a limited number of decimals given by an integer $\nu$. In the approximation algorithm, called $App$, we set $\mu=10^{\nu}$ and we truncate the value of each term below the $\nu$-th decimal. By limiting the value of $\nu$ and thus the precision at which we want to solve the problem, we can maintain the running time of the DP program under control. Moreover, algorithm $App$ provides both a lower and an upper bound on the optimum solution value, as the following proposition shows.
\begin{myProp}[theorem style=plain]{}{Approx}
For the stochastic CNP over trees with unit pair costs and unit deletion costs, algorithm $App$ constitutes an approximation algorithm with time complexity $\O(K^2n^3\mu^2)$ and an approximation bound of $\frac{n(n-1)}{2\mu}$. The truncated objective of the
approximate solution also underestimates the optimal value by at most $\frac{n(n-1)}{2\mu}$.
\end{myProp}
\begin{proof}
The proof is formally the same as the one of Proposition 13 in~\cite{DAM-DCNP}.
\end{proof}

\end{appendices}

\end{document}